\documentclass{amsart}
\usepackage{graphicx,amssymb}
\newtheorem{thm}{Theorem}[section]

\newtheorem{lem}[thm]{Lemma}
\newtheorem{con}[thm]{Conjecture}
\theoremstyle{definition}

\theoremstyle{remark}

\newcommand{\mbold}[1]{\mbox{\boldmath${#1}$}}
\def\beq{\begin{eqnarray}}
\def\eeq{\end{eqnarray}}
\def\d{\mathrm{d}}
\def\Wm{\mathcal{W}_1}
\def\Wc{\mathcal{W}_2}
\begin{document}

\title{The Futures of Bianchi type VII$_0$ cosmologies with vorticity}%
\author[S. Hervik, R.J. van den Hoogen, W.C. Lim, A.A. Coley]{S. Hervik$^{1}$, R.J. van den Hoogen$^{1,2}$, W.C. Lim$^{1}$, A.A. Coley$^{1}$}%
\address{{}$^{1}$Department of Mathematics \& Statistics, Dalhousie University,
Halifax, Nova Scotia,
Canada B3H 3J5 \newline
{}$^{2}$ Department of Mathematics, Statistics \& Computer Science, St. Francis Xavier University, Antigonish, Nova Scotia, Canada B2G 2W5 }%
\email{herviks@mathstat.dal.ca, rvandenh@stfx.ca, wclim@mathstat.dal.ca, \newline aac@mathstat.dal.ca}%

\date{\today}% %
%----------------------------------------------------------------
\begin{abstract} We use expansion-normalised
variables to investigate the Bianchi type VII$_0$ model with a tilted $\gamma$-law
perfect fluid.  We emphasize the late-time asymptotic dynamical behaviour 
of the models and
determine their asymptotic states.  Unlike the other Bianchi models of
solvable type, the type VII$_0$ state space is unbounded.
Consequently we show that,
for a general non-inflationary perfect fluid, one of the curvature variables
diverges at late times, which implies that the type VII$_0$ model
is not asymptotically self-similar to the future.  Regarding the tilt velocity, we show
that for fluids with $\gamma<4/3$ (which includes the important case of
dust, $\gamma=1$) the tilt velocity tends to zero at late times, while for
a radiation fluid, $\gamma=4/3$, the fluid is tilted and its vorticity is
dynamically significant at late times.  For fluids stiffer than radiation
($\gamma>4/3$), the future asymptotic state is an extremely tilted spacetime with
vorticity.  \end{abstract} \maketitle 

\section{Introduction}

Much progess has been made recently in understanding spatially homogeneous
(SH) cosmologies containing a perfect fluid with an equation of state
$p=(\gamma -1)\rho$, where $\gamma$ is a constant.  For the SH cosmologies
known as the Bianchi models, the universe is foliated into space-like
hypersurfaces (defined by the group orbits of the respective model) and the
Einstein equations describe the evolution of these hypersurfaces
\cite{EM,DS1,DS2,BS,rosjan,BN}.  For these models there are two naturally defined
time-like vectors:  the unit vector field, $n^{\mu}$, normal to the group
orbits, and the four-velocity, $u^{\mu}$, of the perfect fluid.  When
analysing the Bianchi models it is common to utilize $n^{\mu}$ as the preferred
timelike vector field in the formalism, while the fluid velocity $u^{\mu}$
may or may not be aligned with $n^{\mu}$.  If $u^{\mu}$ is not aligned with
$n^{\mu}$, the model is called \emph{tilted}, and non-tilted (or orthogonal)
otherwise \cite{KingEllis}.

All of the non-tilted Bianchi models,  except for the type IX model\footnote{However, see for example \cite{HBC,Ringstrom2}.},
have been studied in detail (see e.g.,  \cite{DS1,HWClassB}).  For tilted
Bianchi models the picture is less complete; the type II model was studied in
\cite{HBWII}, the type V model in \cite{Shikin,Collins,CollinsEllis,HWV,Harnett}, 
type VI$_0$
in \cite{hervik,coleyhervik}, types IV and VII$_h$ in \cite{CH,HHC}, a subset of
the type VI$_h$ models in \cite{CH}, and irrotational type VII$_0$ models were studied
in \cite{CH}.  In this paper we will use a dynamical systems approach and
investigate the tilted type VII$_0$ models.

In the class of spatially homogeneous Bianchi models the type VII$_0$
model is a special, and particularly interesting, case.  For Bianchi models
containing a non-tilted perfect fluid, it was shown that the type VII$_0$
model experiences a \emph{self-similarity breaking} at late times
\cite{VII0,VII0rad} (see also \cite{CC1,CC2}).  The irrotational models show the same behaviour
\cite{CH}.  The reason for this self-similarity breaking is because one of
the curvature variables grows unbounded, which leads to an
oscillation of the shear and the curvature variables with a frequency
increasing with time.  We will see that the same behaviour occurs for the
fully tilted type VII$_0$ models; however, in this case there are two
oscillations with different frequencies  (one in the shear and curvature
variables, and the other in the tilt velocity).

\begin{figure}
  \includegraphics[width=7cm]{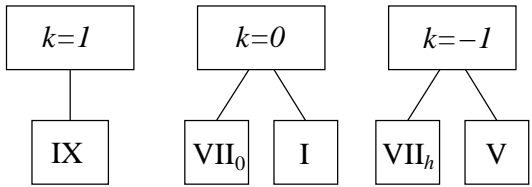}
  \caption{The Bianchi generalisations of the closed ($k=1$), flat ($k=0$) and open ($k=-1$) FRW model.}\label{Fig:FRW}
\end{figure}

\begin{figure}
  \includegraphics[width=7cm]{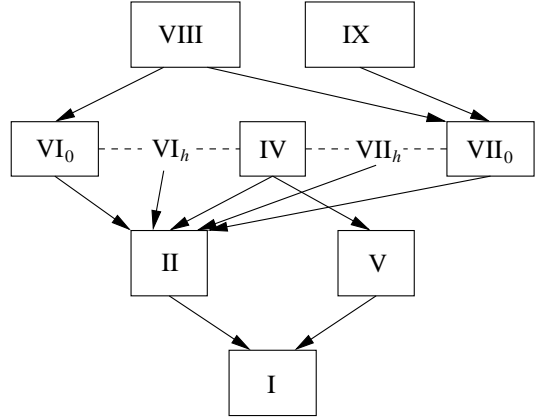}
  \caption{The Bianchi Hierarchy: Solid arrows indicate Lie algebra contractions, or certain special limits of the Lie algebra parameters. Dashed lines indicate one-parameter families of non-isomorphic Lie algebras.  }\label{Fig:Hierarchy}
\end{figure}

The investigation of the type VII$_0$ model is important  for several 
reasons.  First, the type VII$_0$ model is the most general spatially
homogeneous model allowing for the flat Friedmann-Robertson-Walker (FRW) model
as a special case (see Fig.\ref{Fig:FRW}).  Both the Bianchi type I model
and the VII$_0$ generalise the flat FRW model; however, since the Bianchi
type I perfect fluid model does not allow for tilt, only the Bianchi type VII$_0$ model
can be used to investigate the effect of tilt on the evolution
of the universe close to flatness. In addition,
the type VII$_0$ model plays an important role in the Bianchi
hierarchy  (depicted in Fig.\ref{Fig:Hierarchy}).  The
higher up in the hierarchy, the more free parameters the model contains and thus the more
'general' they are.  Fig.  \ref{Fig:Hierarchy} also shows the possible
limits (solid arrows) of the various Lie algebras.  For example, the type
VII$_0$ is a special limit of both the 'most general' type VIII and
type IX models.  In order to understand a particular Bianchi model, a
complete understanding of all its descendants at the lower levels is
needed.  This implies that an understanding of the type VII$_0$ model is necessary
for a full understanding of the two general semi-simple models (i.e., the
type VIII and IX models).

\section{Equations of motion for the tilted VII$_0$ model}
The Bianchi type VII$_0$ model has the Lie algebra of Killing vectors given by:  
\[ 
[{\mbold\xi}_1,{\mbold\xi}_2]={\mbold\xi}_3, \quad [{\mbold\xi}_1,{\mbold\xi}_3]=-{\mbold\xi}_2.
\]
This Lie algebra is solvable and in \cite{CH} a formalism for studying the tilted 
solvable Bianchi models was given. In this paper we will use this formalism to study the Bianchi type VII$_0$ model with full tilt. 

In the dynamical systems approach we introduce expansion-normalised shear variables 
(bold variables are complex variables) $(\Sigma_+, {\mbold\Sigma}_{\times}, {\bold\Sigma}_1)$ 
and the curvature variables $(A,\bar{N},{\bf N}_{\times})$. More specifically, $N^{a}_{~a}=2\sqrt{3}\bar{N}$ while ${\bf N}_{\times}$ consists of the trace-free part of $N_{ab}$. Moreover, for the type VII$_0$ model, $A=0$.  Regarding the 
fluid, $\Omega$ is the expansion-normalised energy density and the tilt velocity is given by 
one real and one complex variable, $v_1$ and ${\bf v}=v_2+iv_3$. We note that the fluid in 
the Class A models has non-zero vorticity if and only if  $N_{ab}v^a\neq 0$ \cite{KingEllis}, 
which implies that the type VII$_0$ model has vorticity\footnote{From \cite{HBWII} 
(see Appendix C), the fluid vorticity
$W_a$ of the Bianchi VII$_0$ model (components with respect to the
$G_3$-adapted frame) is given by
\[
        W_1 = \frac{1}{2B} \frac{1}{1-V^2} N^{CD}v_C v_D v_1
        \ ,\quad
        W_A = \frac{1}{2B} \left[ N_A{}^C v_C + \frac{1}{1-V^2}
                N^{CD}v_C v_D v_A \right]
        \ ,\quad
        W_0 = - v_a W^a\ ,
\]
where the upper case subscripts run from 2 to 3, and
\[B = \frac{1-\frac13(V^2 + \Sigma_{ab} v^a v^b)}{ G_- \sqrt{1-V^2} }\ .
\] 
}
 if and only if ${\bf v}\neq 0$. It is also necessary to introduce the dimensionless time variable, $\tau$, defined by 
\beq 
\frac{\d t}{\d \tau}=\frac{1}{H},
\eeq
where $t$ is the cosmological time and $H$ is the Hubble scalar. 

The equations of motion  in \cite{CH} are given in terms of an arbitrary gauge function 
$\phi$. In terms of the above variables, the gauge transformation is given by 
\beq
\left({\bf N}_{\times},{\mbold\Sigma}_{\times},{\mbold\Sigma}_1,{\bf v}\right)&\mapsto& \left(e^{2i\phi}{\bf N}_{\times},e^{2i\phi}{\mbold\Sigma}_{\times},e^{i\phi}{\mbold\Sigma}_1,e^{i\phi}{\bf v}\right).
 \eeq
Several different gauges were proposed and discussed
(including their advantages and disadvantages). In this paper we will adopt the \emph{'F-gauge'} for 
which $\phi'=0$. This still leaves us with an unspecified \emph{constant} gauge freedom. 
The system of equations will therefore contain one more variable than the number of physical 
degrees of freedom. The physical properties of the system can be extracted by considering 
gauge independent quantites such as, for example, ${\mbold\Sigma}_{\times}{\bf N}^*_{\times}$ and 
${\bf N}^*_{\times}{\bf v}^2$.  

In the 'F-gauge' ($\phi'=0$) the equations of motion are:
\beq \Sigma_+'&=&
(q-2)\Sigma_++{3}|{\mbold\Sigma}_1|^2-2|{\bf N}_{\times}|^2
+\frac{\gamma\Omega}{2G_+}\left(-2v_1^2+|{\bf v}|^2\right) \label{eq:Sigma+eq}\\
{\mbold\Sigma}_{\times}'&=&
(q-2){\mbold\Sigma}_{\times}+\sqrt{3}{\mbold\Sigma}_{1}^2
-2\sqrt{3}{\bf N}_{\times}\bar{N}
+\frac{\sqrt{3}\gamma\Omega}{2G_+}{\bf v}^2
\\
{\mbold\Sigma}'_{1}&=& \left(q-2-3\Sigma_+\right){\mbold
\Sigma}_{1} -\sqrt{3}{\mbold\Sigma}_{\times}{\mbold\Sigma}_{1}^*
+\frac{\sqrt{3}\gamma\Omega v_1}{G_+}{\bf v}
\\
 {\bf N}_{\times}'&=& \left(q+2\Sigma_+\right){\bf N}_{\times}+2\sqrt{3}{\mbold\Sigma}_{\times}\bar{N}\\
\bar{N}'&=&
\left(q+2\Sigma_+\right)\bar{N}+2\sqrt{3}\mathrm{Re}\left({\mbold\Sigma}_{\times}^*{\bf
N}_{\times}\right)
 \label{eq:Aeq}\eeq
The equations for the fluid are
\beq
\quad \Omega'&=& \frac{\Omega}{G_+}\Big\{2q-(3\gamma-2)
 +\left[2q(\gamma-1)-(2-\gamma)-\gamma\mathcal{S}\right]V^2\Big\}\label{eq:Omega}
 \quad \\
 v_1' &=& \left(T+2\Sigma_+\right)v_1-2\sqrt{3}\mathrm{Re}\left({\mbold\Sigma}_{1}{\bf v}^*\right) +\sqrt{3}\mathrm{Im}({\bf N}_{\times}^*{\bf v}^2)\\
 {\bf v}'&=& \left(T-\Sigma_+-i\sqrt{3}\bar{N}v_1\right){\bf v}-\sqrt{3}\left({\mbold\Sigma}_{\times}+i{\bf N}_{\times}v_1\right){\bf v}^* \\
  V'&=&
\frac{V(1-V^2)}{1-(\gamma-1)V^2}\left[(3\gamma-4)-\mathcal{S}\right]
\eeq where
\beq q&=& 2\Sigma^2+\frac
12\frac{(3\gamma-2)+(2-\gamma)V^2}{1+(\gamma-1)V^2}\Omega\nonumber \\
\Sigma^2 &=& \Sigma_+^2+|{\mbold{\Sigma}}_{\times}|^2+|{\mbold\Sigma}_{1}|^2\nonumber \\
\mathcal{S} &=& \Sigma_{ab}c^ac^b, \quad c^ac_{a}=1, \quad v^a=Vc^a,\quad \nonumber \\
 V^2 &=& v_1^2+|{\bf v}|^2,\quad  \nonumber \\
G_+ &=& 1+(\gamma-1)V^2, \nonumber \\
 T&=& \frac{(3\gamma-4)(1-V^2)+(2-\gamma)V^2\mathcal{S}}{1-(\gamma-1)V^2}.
\label{eq:defs}
\eeq
These variables are subject to the constraints
\beq
1&=& \Sigma^2+|{\bf N}_{\times}|^2+\Omega \label{const:H}\\
0 &=& 2\mathrm{Im}({\mbold\Sigma}_{\times}^*{\bf N}_{\times})+\frac{\gamma\Omega v_1}{G_+} \label{const:v1}\\
0 &=&
i\mbold{\Sigma}_{1}\bar{N}+i{\mbold{\Sigma}}^*_{1}{\bf
N}_{\times}+\frac{\gamma\Omega {\bf v}}{G_+} \label{const:v2}
\eeq 
The
parameter $\gamma$ will be assumed to be in the interval $\gamma\in ( 0,2)$.  The generalised
Friedmann equation, eq.(\ref{const:H}), yields an expression which effectively determines the
energy density $\Omega$.  The state vector will therefore be considered to be ${\sf
X}=[\Sigma_+,{\mbold\Sigma}_{\times},{\mbold\Sigma}_1,{\bf N}_{\times},\bar{N},v_1,{\bf v}]$
modulo the constraint equations (\ref{const:v1}) and (\ref{const:v2}).  Thus the real dimension
of the dynamical system is eight of which one is the remaining gauge freedom; i.e., the dimension
of the \emph{physical} state space is seven.

The dynamical system is invariant under the following discrete symmetries: 
\beq
\phi_1: && [\Sigma_+,{\mbold\Sigma}_{\times},{\mbold\Sigma}_1,{\bf N}_{\times},\bar{N},v_1,{\bf v}]\mapsto [\Sigma_+,{\mbold\Sigma}_{\times},{\mbold\Sigma}_1,-{\bf N}_{\times},-\bar{N},-v_1,-{\bf v}]\nonumber \\
\phi_2: && [\Sigma_+,{\mbold\Sigma}_{\times},{\mbold\Sigma}_1,{\bf N}_{\times},\bar{N},v_1,{\bf v}]\mapsto [\Sigma_+,{\sf u}^2{\mbold\Sigma}_{\times}^*,-{\sf u}{\mbold\Sigma}_1^*,{\sf u}^2{\bf N}_{\times}^*,\bar{N},-v_1,{\sf u}{\bf v}^*], ~~{\sf u}\in S^1\nonumber \\ 
\phi_3: && [\Sigma_+,{\mbold\Sigma}_{\times},{\mbold\Sigma}_1,{\bf N}_{\times},\bar{N},v_1,{\bf v}]\mapsto [\Sigma_+,{\mbold\Sigma}_{\times},-{\mbold\Sigma}_1,{\bf N}_{\times},\bar{N},v_1,-{\bf v}]\nonumber.
\eeq
These discrete symmetries imply that without loss of generality we can restrict the 
variables $\bar{N}\geq 0$ and $v_1\geq 0$. \footnote{There is a sublety regarding this 
choice since, in general, $v_1=0$  is not an invariant subspace. The state space can be 
considered an orbifold with a mirror symmetry at $v_1=0$; in particular, this means that 
any equilibrium point in the region $v_1>0$ has an  analogous equilibrium point in the 
region $v_1<0$.}  We also note that the free parameter ${\sf u}$ in $\phi_2$ is, in fact, 
the remaining gauge transformation. 

\subsection{Invariant Subspaces}In this analysis we will be concerned with the following invariant sets:
\begin{enumerate}
\item{} $T(VII_0)$: The general tilted type VII$_0$ model with $\bar{N}^2-\left|{\bf N}_{\times}\right|^2>0$.
\item{} $T_1(VII_0)$: The irrotational type VII$_0$ models defined by $\bar{N}^2-\left|{\bf N}_{\times}\right|^2>0$, ${\mbold\Sigma_1}={\bf v}=0$. 
\item{} $F(VII_0)$: The set of fixed points of $\phi_2$ (without loss of generality we 
can set ${\sf u}=1$) defined by $v_1=\mathrm{Im}({\mbold\Sigma}_{\times})=\mathrm{Re}({\mbold\Sigma}_1)=\mathrm{Im}({\bf N}_{\times})=\mathrm{Im}({\bf v})=0$.
\item{} $B(VII_0)$: Non-tilted Bianchi type VII$_0$ models with $\bar{N}^2-\left|{\bf N}_{\times}\right|^2>0$, $v_1={\bf v}=0$.
\item{} $T(II)$: The general tilted type II model given by $\bar{N}^2-\left|{\bf N}_{\times}\right|^2=0$.
\item{} $B(I)$: Type I: $\bar{N}={\bf N}_{\times}=V=0$.
\item{} $\partial T(I)$: ``Tilted'' vacuum type I: $\Omega=\bar{N}={\bf N}_{\times}=0$.
\end{enumerate}
We note that the closure of the set $T(VII_0)$ is given by:
\beq
\overline{T(VII_0)}=T(VII_0)\cup T(II)\cup B(I)\cup \partial T(I).
\eeq
Morever, in $\overline{T(VII_0)}$ we have the bounds
\beq
\Sigma_+^2+|{\mbold{\Sigma}}_{\times}|^2+|{\mbold\Sigma}_{1}|^2+\left|{\bf N}_{\times}\right|^2\leq 1, \quad v^2_1+|{\bf v}|^2\leq 1;
\eeq
hence, all variables are bounded except for $\bar{N}$ (which can become arbitrary large). 
\subsection{Monotonic functions}
There are three monotonic functions which are useful for our analysis:
\beq
Z_1& \equiv &\alpha\Omega^{1-\Gamma}, \quad \alpha=\frac{(1-V^2)^{\frac 12(2-\gamma)}}{G_+^{1-\Gamma}V^\Gamma}, \quad \Gamma=\frac 67\gamma, \\
Z_1' &=& \left[2(1-\Gamma)q+(2+2\Gamma-3\gamma)+\Gamma\mathcal{S}\right]Z_1.\nonumber 
\eeq
We note that (by the same trick as in \cite{CH})
\beq 
&&2(1-\Gamma)q+(2+2\Gamma-3\gamma)+\Gamma\mathcal{S} \nonumber \\
&& \geq \frac 17(14-15\gamma)(2\Sigma^2+|{\bf N}_{\times}|^2)+\frac{\gamma\left[18(1-\gamma)+(10-9\gamma)V^2\right]}{7G_+}\Omega.\nonumber
\eeq
Thus $Z_1$ is monotonically increasing in $T(VII_0)$ for $\gamma\leq 14/15$. 

\beq
Z_2&\equiv & \frac{\left(\bar{N}^2-|{\bf N}_{\times}|^2\right)^{n}\beta\Omega}{\left(1+n\Sigma_+\right)^{2(1+n)}},\quad \beta\equiv\frac{(1-V^2)^{\frac12(2-\gamma)}}{G_+}, \quad n\equiv\frac 14(3\gamma-2) \\
 Z_2'&=&\frac{4}{1+n\Sigma_+}\Bigg\{\left(\Sigma_++n\right)^2+(1-n^2)|{\mbold\Sigma}_{\times}|^2 
 +\frac{(1+n)(2-5n)}{2}|{\mbold\Sigma}_1|^2\nonumber \\
&&\qquad +\frac{(1+2n)(1+n)\Omega}{6G_+}\left[2(1-n)v_1^2+(2-5n)|{\bf v}|^2\right]\Bigg\}Z_2.\nonumber 
\eeq 
$Z_2$ is monotonically increasing for the following subspaces:
\footnote{We note that all of these functions are also monotonic for the VI$_0$ model, 
where the importance of the value $\gamma=6/5$ is more clear. In fact, using these monotonic 
functions we can prove that the local attractors found in \cite{hervik} are, indeed, global attractors.} $T_1(VII_0)$ for $0<\gamma<2$, and  $T(VII_0)$ for $0<\gamma\leq 6/5 $.

\beq
Z_3 &\equiv&\frac{\left[\bar{N}|{\bf v}|^2+\mathrm{Re}({\bf N}_{\times}^*{\bf v}^2)\right]^2}{(1-V^2)^{2(2-\gamma)}\beta\Omega}, \\
Z_3' &=&3(5\gamma-6) Z_3. \nonumber
\eeq
Thus $Z_3$ is monotonically decreasing ($\gamma <6/5$) or increasing ($6/5<\gamma$) in $T(VII_0)\setminus T_1(VII_0)$.   

\section{Qualitative analysis}
The above monotonic functions allow us to obtain some results regarding the asymptotic behaviour of Bianchi type VII$_0$ universes: 
\begin{thm} 
For $0<\gamma<6/5$, all tilted Bianchi models  (with $\Omega>0$, $V<1$) of type VII$_0$ are asymptotically non-tilted at late times. 
\end{thm}
\begin{proof} For $0<\gamma\leq 14/15$ we can use the monotonic function $Z_1$, which 
immediately gives the desired result. For $2/3<\gamma<6/5$, and using the monotonic 
function $Z_3$, we get $\bar{N}|{\bf v}|^2+
\mathrm{Re}({\bf N}_{\times}^*{\bf v}^2)\rightarrow 0$. 
Moreover, using $Z_2$, we get $\bar{N}\rightarrow \infty$ 
and hence $|{\bf v}|\rightarrow 0$. This implies that the solution  asymptotically
approaches the invariant subspace $T_1(VII_0)$. 
The $T_1(VII_0)$ analysis \cite{CH} can now be applied which shows that $v_1\rightarrow 0$. 
The theorem now follows. 
\end{proof}
In fact, we believe that the type VII$_0$ models are asymptotically non-tilted for $0<\gamma<4/3$; however, this is not covered by this theorem. 

Another important observation is the divergence of the variable $\bar{N}$ at late times. In fact, we firmly believe that: 
\begin{con}
For a non-inflationary perfect fluid ($2/3<\gamma<2$) and any initial condition in $T(VII_0)$ with $V<1$ and $\Omega>0$, we have that 
\[ \lim_{\tau\rightarrow +\infty}\left|\bar{N}\right|=\infty.\]
\label{conjecture}\end{con}
There are several results that support this conjecture.  First, use of the monotonic
function $Z_2$ immediately establishes this result for $T_1(VII_0)$ and for $2/3<\gamma<6/5$
with general tilt (which includes the important case $\gamma=1$).  Moreover, using $Z_3$ we have
that for $6/5<\gamma$, $\bar{N}\rightarrow \infty$ or $V\rightarrow 1$.  Second, a ``local
analysis'' shows that the conjecture is true locally.  And third, in our numerical analysis
no other behaviour has been seen.  In Appendix \ref{app:num} some of the numerical plots
are presented; in
particular, we see that $M\equiv 1/\bar{N}\rightarrow 0$ for various values of $\gamma$.

\subsection{Equilibrium points} 
\subsubsection{$B(I)$: equilibrium points of Bianchi type I} 
\begin{enumerate}
\item{} $\mathcal{I}(I)$: $q=\tfrac{1}2(3\gamma-2)$, $\Omega=1$, $\Sigma_+={\mbold\Sigma}_{\times}={\mbold\Sigma}_{1}={\bf N}_{\times}=\bar{N}=V=0$\\
 Eigenvalues: $\tfrac{1}2(3\gamma-2)[\times 3]$, $-\tfrac{1}2(2-\gamma)[\times 4]$.\\
 The remaining equilibrium points of type I are all in $\partial T(I)$. 
\end{enumerate}
\subsubsection{$T(II)$: equilibrium points of Bianchi type II} All of the equilibrium points in the set $T(II)$ are unstable. They are all given in \cite{HHC} (in 'F-gauge').
\subsubsection{$T(VII_0)$: equilibrium points of Bianchi type VII$_0$} 
Using the monotonic functions
$Z_3$ and $Z_2$, it is possible to show that there are no equilibrium points
in $T(VII_0)$, apart from the following:
 \begin{enumerate}
 \item{} $\mathcal{L}^\alpha_k(VII_0)$: $q=2$, $\Sigma_{+}=-1$, $\bar{N}=k=\text{constant}$, ${\mbold\Sigma}_{\times}={\mbold\Sigma}_{1}={\bf N}_{\times}=\Omega=0$, with the tilt velocities given by:
 \begin{enumerate}
 \item{} $0<\gamma<2$, $v_1={\bf v}=0$.
 \item{} $0<\gamma<2$, $v_1=1$, ${\bf v}=0$.
 \item{} $0<\gamma<2$, $v_1=0$, $|{\bf v}|=1$.
 \item{} $\gamma=1$, $v_1=0$, $0<|{\bf v}|<1$.
 \end{enumerate}
 \item{} $\mathcal{F}_k(VII_0)$: $q=0$, $\bar{N}=k=\text{constant}$, $\Omega=1$, $\Sigma_+={\mbold\Sigma}_{\times}={\mbold\Sigma}_{1}={\bf N}_{\times}=V=0$, $\gamma=\frac 23$.\\
 Eigenvalues: $0$, $-2[\times 2]$, $-1\pm\sqrt{1-12k^2}[\times 2]$. \\
 This is an attracting set (for $\gamma=2/3$). 
 \end{enumerate}
All of the equilibrium points in
$\overline{T(VII_0)}$, including the ones in the sets $T(II)$, $B(I)$, $\partial T(I)$, can be
shown to be unstable into the future for $2/3<\gamma<2$.

\subsection{Late time analysis} 
For the type VII$_0$ model it is convenient to solve constraint (\ref{const:v2}) to obtain an expression for ${\mbold\Sigma}_1$:
\beq
{\mbold\Sigma}_1=\frac{i\gamma\Omega}{G_+\left(\bar{N}^2-\left|{\bf N}_{\times}\right|^2\right)}\left(\bar{N}{\bf v}+{\bf N}_{\times}{\bf v}^*\right).
\eeq

In the following we will introduce the variable 
\beq
M\equiv \frac{1}{\bar{N}}. 
\eeq
In light of conjecture \ref{conjecture}, we have that $M\rightarrow 0$. In particular, 
\beq
{\mbold\Sigma}_1=\frac{i\gamma M\Omega}{G_+\left(1-M^2\left|{\bf N}_{\times}\right|^2\right)}\left({\bf v}+M{\bf N}_{\times}{\bf v}^*\right)=M{\bf B},
\eeq
where ${\bf B}$ is a bounded (complex) function. 

Due to the oscillatory behaviour of the system, we introduce the following variables
\beq
{\mbold\Sigma}_{\times}+i{\bf N}_{\times} &\equiv & e^{2i\psi}{\bf X}, \\
{\mbold\Sigma}_{\times}-i{\bf N}_{\times} &\equiv & e^{-2i\psi}{\bf Y}, \\
{\bf v} &\equiv & e^{-i\theta}v_2,
\eeq
where 
\beq
\psi' &=& \frac{\sqrt{3}}{M}, \\
\theta' &=& \frac{\sqrt{3}}{M}\left\{v_1+\frac 12M\mathrm{Im}\left[(1+v_1){\bf X}e^{2i(\theta+\psi)}+ (1-v_1){\bf Y}e^{2i(\theta-\psi)}\right]\right\}.
\eeq
The angular variables $\psi$ and $\theta$ are introduced to take care of the rapid oscillation 
as $M\rightarrow 0$. We note that the variables $\psi$ and $\theta$ are not in 
synchronization since $v_1<1$. Hence, in general, we expect two different oscillations with different frequencies.  
The equation of motions for these variables are given in Appendix \ref{App:A}. In this case we can 
explicitly see how the oscillatory terms enter into the equations of motion. Moreover, 
we note that both of these rapid oscillations are observable; e.g., by considering the scalars: 
\beq
S_1&=&({\mbold\Sigma}_{\times}+i{\bf N}_{\times})({\mbold\Sigma}_{\times}^*+i{\bf N}_{\times}^*)=e^{4i\psi}{\bf X}{\bf Y}^*, \nonumber \\
S_2&=&({\mbold\Sigma}_{\times}^2+{\bf N}_{\times}^2)({\bf v}^*)^4=e^{4i\theta}{\bf X}{\bf Y}v_2^4. \nonumber
\eeq

\subsection{Reduced System} \label{sect:redsys}
The idea is that as $M\rightarrow 0$ at late times, the system of equations effectively reduces to a much simpler system of equations. In Appendix \ref{App:A} the idea behind this reduction is explained in more detail; by introducing new variables this reduction of the full system is manifest as $M\rightarrow 0$. This method was also used in the previous analyses of the type VII$_0$ model \cite{VII0,VII0rad,CH} with success. Moreover, in Appendix \ref{App:B} a linear analysis is given to put bounds on solutions of the reduced system with respect to the full system. 

Therefore we assume that $M\rightarrow 0$, which implies that ${\mbold\Sigma}_1\rightarrow 0$. Consequently, from above we have that \emph{all oscillatory terms effectively cancel}. Defining 
\beq
\sigma_1&\equiv& \left|{\bf N}_{\times}\right|^2+ \left|{\mbold\Sigma}_{\times}\right|^2=\frac{1}{2}\left(|{\bf X}|^2+|{\bf Y}|^2\right),\\
\sigma_2&\equiv& 2\mathrm{Im}\left({\mbold\Sigma}_{\times}{\bf N}_{\times}^*\right)=-\frac{1}{2}\left(|{\bf X}|^2-|{\bf Y}|^2\right),
\eeq
the system given by eqs.(\ref{NewSigma+}-\ref{Newv2}) effectively reduces to the following system: 
\beq
\Sigma_+'&=&
(Q-2)\Sigma_+-\sigma_1
+\frac{\gamma\Omega}{2G_+}\left(-2v_1^2+v_2^2\right) \\
\sigma_1'&=& 2(Q+\Sigma_+-1)\sigma_1\\ 
M'&=&
-\left(Q+2\Sigma_+\right)M\\
\Omega'&=& \frac{\Omega}{G_+}\Big\{2Q-(3\gamma-2)
 +\left[2Q(\gamma-1)-(2-\gamma)-\gamma\mathcal{S}\right]V^2\Big\}
 \\
v_1' &=& \left(T+2\Sigma_+\right)v_1\\
v_2'&=&(T-\Sigma_+)v_2
\eeq
where
\beq
Q&=& 2\Sigma^2_++\sigma_1+\frac
12\frac{(3\gamma-2)+(2-\gamma)V^2}{1+(\gamma-1)V^2}\Omega, \\
V^2\mathcal{S} &=& \left(-2v_1^2+v_2^2\right)\Sigma_+.
\eeq
These variables are subject to the constraint
\beq
1&=& \Sigma^2_++\sigma_1+\Omega.
\label{eq:redHamiltonian}\eeq
Furthermore, $\sigma_2$ is determined from
\[ 
\sigma_2=-\frac{\gamma\Omega v_1}{G_+},
\]
which gives the bound 
\beq 
\sigma_1\geq \frac{\gamma\Omega |v_1|}{G_+}.
\eeq
Using the constraint (\ref{eq:redHamiltonian}) we can solve for $\sigma_1$ or $\Omega$. 

\subsubsection{Monotonic functions for the reduced system}
By the same trick as in Appendix \ref{App:A}, monotonic functions of the reduced system will at sufficiently late times also be monotonic for the full system. It is therefore useful to list some of the monotonic functions for the reduced system. 
\beq
R_1&\equiv & \frac{\beta\Omega}{\sigma_1^m(1-m\Sigma_+)^{2(1-m)}}, \quad m=\frac{1}{2}(4-3\gamma), \nonumber\\
R_1'&=& \frac{2}{1-m\Sigma_+}\left[(\Sigma_+-m)^2+\frac{m(1-m)(2-m)|{\bf v}|^2\Omega}{G_+}\right]R_1,\nonumber 
\eeq
which is monotonically increasing for $2/3<\gamma<4/3$ if ${\bf v}\neq 0$, and for $2/3<\gamma<2$ if ${\bf v}=0$. 

\beq
R_2 &\equiv & \frac{v_1|{\bf v}|^2}{(1-V^2)^{\frac 32(2-\gamma)}}, \nonumber \\ 
R_2' &=& 3(3\gamma-4)R_2,\nonumber 
\eeq
which is monotonically decreasing or increasing for $\gamma<4/3$ and $\gamma>4/3$, respectively.

 \beq
 R_3 &\equiv & \beta\Omega, \nonumber \\
 R_3'&=& \left[2\Sigma_+^2+\frac{(4-3\gamma)}{G_+}\left[(1-V^2)(1-\Omega)+ \gamma V^2\right]\right]R_3,
 \eeq
which is monotonically increasing for $\gamma\leq  4/3$.
\subsubsection{Equilibrium points for the reduced system}
\paragraph{} $\widetilde{P_1}$: $\Sigma_+=\sigma_1=M=v_1=v_2=0$, $\Omega=1$, $Q=\frac 12(3\gamma-2)$, $2/3<\gamma<2$. 
\newline Eigenvalues: 
\[ -\frac 12(3\gamma-2),\quad -(4-3\gamma)[\times 3],\quad -\frac 32(2-\gamma). \]
\paragraph{} $\widetilde{P_2}$: $\Sigma_+=-\frac 12(3\gamma-4)$, $\sigma_1=\frac{3(2-\gamma)[(3\gamma-4)-v_1^2(5\gamma-4)]}{4(1-v_1^2)}$, $M=v_2=0$, $\Omega=\frac{3(2-\gamma)G_+}{2(1-v_1^2)}$, $Q=\frac 12(3\gamma-2)$, $0\leq v_1\leq \frac{3\gamma-4}{5\gamma-4}$, $4/3<\gamma<2$. 
\newline Eigenvalues:
\[ 0,\quad -3(4-3\gamma), \quad -\frac 32(2-\gamma),\quad -\frac{(2-\gamma)}{4}\left[1\pm\sqrt{1-\frac{4[(3\gamma-4)+v_1^2(5\gamma-4)]}{1-(\gamma-1)v_1^2}}\right].\] 
\paragraph{} $\widetilde{P_3}$: $\Sigma_+=M=0$, $\sigma_1=\frac{2(v_2^2-2v_1^2)}{3(1-v_1^2+v_2^2)}$, $\Omega=\frac{3+v_1^2+v_2^2}{3(1-v_1^2+v_2^2)}$, $Q=1$, $\gamma=\frac 43$, $0\leq (2+3v_1)v_1\leq v_1^2+v_2^2<1$.  
\newline Eigenvalues: 
\[ 0[\times 2], \quad -1, \quad -\frac 12\left(1\pm\sqrt{1-\frac{16\left[2v_1^2(1-v_1^2)+v_2^2(2-v_2^2)\right]}{(1-v_1^2+v_2^2)(3-V^2)}}\right). \]
\paragraph{} $\widetilde{P_4}$: $\Sigma_+=M=0$, $\sigma_1=\frac{1-3v_1^2}{3(1-v_1^2)}$, $v_2=\sqrt{1-v_1^2}$, $\Omega=\frac{2}{3(1-v_1^2)}$, $Q=1$, $0\leq v_1\leq 1/3$, $2/3<\gamma<2$. 
\newline Eigenvalues: 
\[ 0,\quad -\frac{2(3\gamma-4)}{2-\gamma},\quad -1, \quad -\frac 12\left(1\pm i\sqrt{3+12v_1^2}\right).\]

\section{Late time behaviour} 

\begin{thm} 
For all tilted Bianchi models of type VII$_0$ with a perfect fluid stiffer than 
radiation ($4/3<\gamma<2$) and where the fluid has non-zero vorticity (${\bf v}\neq 0$), 
the fluid will asymptotically approach a state of extreme tilt; i.e., $\lim_{\tau\rightarrow\infty}V=1$. 
\end{thm}
\begin{proof} 
From the monotonic function $Z_3$ we have that $V\rightarrow 1$ or $\bar{N}\rightarrow\infty$. 
First, we assume that $\bar{N}\rightarrow\infty$. At sufficiently late times we can then
use the reduced system. For $v_1\neq 0$, we use the function $R_2$ which is monotonically 
increasing at sufficiently late times. This shows that $V\rightarrow 1$. We assume, therefore, that $v_1=0$. For $v_1=0$ there is a monotonic function given by 
\[ R_4=\frac{\sigma_1|{\bf v}|^2}{(1-V^2)^{(2-\gamma)}}, \quad R_4'=2\left[(3\gamma-4)\left(1+\frac{1-V^2}{2G_+}\Omega\right)+\Sigma_+^2\right]R_4,\]
which again implies $V\rightarrow 1$. The theorem now follows. 
\end{proof} 
As summarized in Table \ref{tab:outline}, the tilt becomes extreme only in the case 
$\frac43 < \gamma < 2$ for fluid with vorticity.
In the case of zero vorticity, a very different tilt behaviour is
observed; namely, the tilt tends to a non-extreme limit, as described by
the equilibrium point $\widetilde{P}_2$ \cite{CH,LDW}.

\begin{thm}
Consider a perfect fluid Bianchi type VII$_0$ model with $2/3<\gamma\leq 4/3$ for which $\bar{N}\rightarrow \infty$. If $V<1$, then:   
\begin{enumerate}
\item{} $\lim_{\tau\rightarrow\infty}(\Sigma_+,V,\sigma_1)=(0,0,0)$ for $2/3<\gamma<4/3$, 
\item{} $\lim_{\tau\rightarrow\infty}\Sigma_+=0$ for $\gamma=4/3$.
\end{enumerate}
\end{thm} 
\begin{proof}
Use $R_3$. 
\end{proof}
\subsection{Decay rates for the case $\frac23 < \gamma < \frac43$}
By assuming an ansatz with coefficients and exponents to be determined, we obtain the decay rates when $\frac23 < \gamma < \frac43$ as follows:
\begin{align}
	\Sigma_+ &\approx 
	- \frac{2\hat{\sigma}_1}{3\gamma-2} e^{-(4-3\gamma)\tau}
	+ \hat{\Sigma}_+ e^{-\frac32(2-\gamma)\tau},
\label{Sp_expansion}
\\
	\sigma_1 &\approx \hat{\sigma}_1 e^{-(4-3\gamma)\tau},
\\
	v_1 &\approx \hat{v}_1 e^{-(4-3\gamma)\tau},
\\
        v_2 &\approx \hat{v}_2 e^{-(4-3\gamma)\tau},
\\
	M &\approx \hat{M} e^{-\frac12(3\gamma-2)\tau}.
\end{align}
The angular variables are given asymptotically by:
\begin{align}
\psi &\approx \hat{\psi}+\frac{2\sqrt{3}}{(3\gamma-2)\hat{M}}e^{\frac12(3\gamma-2)\tau}, \\
\theta &\approx
\begin{cases}  \hat{\theta}, & 2/3<\gamma<10/9 \\
\hat{\theta}+\frac{\sqrt{3}\hat{v}_1}{\hat{M}}\tau, & \gamma=10/9 \\
\hat{\theta}+\frac{2\sqrt{3}\hat{v}_1}{(9\gamma-10)\hat{M}}e^{\frac12(9\gamma-10)\tau}, & 10/9<\gamma<4/3.
\end{cases}
\end{align}
The meaning of the bifurcation value $\gamma=10/9$ is more clear if we calculate the fluid vorticity; in particular, to leading order we have:
\beq
W^aW_a\propto e^{(9\gamma-10)\tau}, \qquad W^0W_0\propto e^{3(5\gamma-6)\tau}.
\eeq
We note that the bifurcation values for the vorticity and tilt at $\gamma=10/9$ and $\gamma=4/3$, respectively, coincide with with the values found in \cite{BarrowTipler}.  

The Hubble-normalised Weyl scalars $\Wm$ and $\Wc$, defined by
\begin{equation}
	\Wm = \frac{C_{abcd}C^{abcd}}{48H^4}\, ,\quad
	\Wc = \frac{C_{abcd}{}^*C^{abcd}}{48H^4}\, ,
\end{equation}
evolve as
\beq\Wm + i \Wc &\approx &- \frac{12}{M^2} S_1 \nonumber \\
	 \Rightarrow \quad 
	(\Wm,\Wc) &\approx & - \frac{12}{\hat{M}^2}
\sqrt{\hat{\sigma}_1^2-\gamma^2 \hat{v}_1^2} \, e^{6(\gamma-1)\tau}
	(\cos 2\psi_2,\sin 2\psi_2)\ ,
\eeq
where $ 0 \leq \hat{v}_1 \leq \frac{\hat{\sigma}_1}{\gamma}$, $\hat{M}$ is the 
constant from the variable $M$, and $\psi_2$ is defined by $S_1 = |S_1| e^{2i\psi_2}$. 
The angular variable $\psi_2$ is related asymptotically to the 
variable $\psi$ via $\psi_2\approx 2\psi+\psi_0$, where $\psi_0$ is a constant. 

We note that the first term in (\ref{Sp_expansion}) is dominant, and the
second term is included to display the constant $\hat{\Sigma}_+$.
The Weyl scalars diverge for $1 \leq \gamma < \frac43$. In the case
$\gamma =1$, the magnitude of the Weyl scalars are asymptotically constant.

The decay rates are verified by numerical simulations. 

\subsection{Decay rates for the radiation case: $\gamma = \frac43$}

The quantitative late-time dynamics 
for the case of radiation ($\gamma=\frac43$) is of particular interest.
The reduced system in this case is
\begin{align}
	\Sigma_+' &= - \Sigma_+ + \Sigma_+^3 - \sigma_1
	+ \frac{2}{3+v_1^2 + v_2^2}(1-\Sigma_+^2-\sigma_1)(-2v_1^2+v_2^2)
\\
	\sigma_1' &= 2(1+\Sigma_+)\Sigma_+ \sigma_1
\\
	v_1' &= \frac{6(1-v_1^2)}{3-v_1^2-v_2^2}\Sigma_+ v_1
\\
	v_2' &= -\frac{3(1+v_1^2-v_2^2)}{3-v_1^2-v_2^2}\Sigma_+ v_2\ ,
\end{align}
with the bound
\begin{equation}
\label{my_bound}
	\sigma_1 \geq \frac{4(1-\Sigma_+^2-\sigma_1)}{3+v_1^2+v_2^2}|v_1|\ .
\end{equation}
In view of the limit $\widetilde{P}_3$, we assume the ansatz
\begin{align}
	\Sigma_+ &= \bar{\Sigma}_+ e^{-k\tau}
\\
	\sigma_1 &= \hat{\sigma}_1 + \bar{\sigma}_1 e^{-k\tau}
\\
	v_1 &= \hat{v}_1 + \bar{v}_1 e^{-k\tau}
\\
        v_2 &= \hat{v}_2 + \bar{v}_2 e^{-k\tau} \ ,
\end{align}
where $k$, the hatted and the barred variables are constants.
The ansatz is substituted into the evolution equations and terms with
equal power are matched. We determine the following constants:
\begin{align}
	\hat{\sigma}_1 &= \frac{2(\hat{v}_2^2 - 2\hat{v}_1^2)}{
	3(1-\hat{v}_1^2+\hat{v}_2^2)}
\\
	\bar{\sigma}_1 &= -\frac{2\bar{\Sigma}_+ \hat{\sigma}_1}{k}
\\
	\bar{v}_1 &= -\frac{1}{k} \frac{6(1-\hat{v}_1^2)}{
	3-\hat{v}_1^2-\hat{v}_2^2}\bar{\Sigma}_+ \hat{v}_1
\\
	\bar{v}_2 &= \frac{1}{k} \frac{3(1-\hat{v}_1^2-\hat{v}_2^2)}{
	3-\hat{v}_1^2-\hat{v}_2^2}\bar{\Sigma}_+ \hat{v}_2
\\
	k_{\pm} &= \frac12\left(1 \pm \sqrt{1-
	\frac{16[2\hat{v}_1^2(1-\hat{v}_1^2) + \hat{v}_2^2(2-\hat{v}_2^2)]
	}{(1-\hat{v}_1^2+\hat{v}_2^2)(3-\hat{v}_1^2-\hat{v}_2^2)}
	} \, \right)\ .
\end{align}
That there are two solutions for $k$ means that the ansatz should have
been extended to include two modes. The remaining arbitrary constants are
$\hat{v}_1$, $\hat{v}_2$ and two $\bar{\Sigma}_{+(\pm)}$'s.
The bound (\ref{my_bound}) restricts $\hat{v}_1$ and $\hat{v}_2$ as
follows:
\begin{equation}
	2 \hat{v}_1 (1+\hat{v}_1) \leq \hat{v}_2^2 \leq 1-\hat{v}_1^2
	\ ,\quad
	0 \leq \hat{v}_1 \leq \tfrac13 \ .
\end{equation}
 We note that when the $k_\pm$ are complex, the
late-time approach is oscillatory.

The angular variables are given asymptotically by:
\begin{align}
\psi &\approx \hat{\psi}+\frac{\sqrt{3}}{\hat{M}}e^{\tau}, \\
\theta &\approx  \hat{\theta}+ \frac{\sqrt{3}\hat{v}_1}{\hat{M}}e^{\tau}.
\end{align}

The Weyl scalars 
diverge as
\begin{equation}
	(\Wm,\Wc) \approx -\frac{8}{\hat{M}^2} 
	\frac{\sqrt{ (\hat{v}_2^2-2\hat{v}_1^2)^2 -4\hat{v}_1^2}
	}{(1-\hat{v}_1^2 + \hat{v}_2^2)} \, e^{2\tau} 
	(\cos 2\psi_2,\sin 2\psi_2)\, ,
\end{equation}
where $\hat{M}$ is the constant from $M \approx \hat{M} e^{-\tau}$, and
$\psi_2$ is again defined by $S_1 = |S_1| e^{2i\psi_2}$.

\subsection{Decay rates for the case $\frac43 < \gamma < 2$}

\begin{align}   
        M &\approx \hat{M} e^{-\tau}
\\
	1-V^2 &\approx C_{(1-V^2)} e^{-\frac{2(3\gamma-4)}{2-\gamma}\tau}
\end{align}
and $\Sigma_+$, $\sigma_1$, $v_1$ and $v_2$ have the form
\begin{equation}
	y \approx \hat{y} + C_1 e^{-\frac{2(3\gamma-4)}{2-\gamma}\tau} 
	+ C_2 e^{-\frac12\tau}\cos\left(\sqrt{3+12\hat{v}_1^2}\,
		\tau\right)
	+ C_3 e^{-\frac12\tau}\sin\left(\sqrt{3+12\hat{v}_1^2}\,
		\tau\right) \ ,
\end{equation}
with
\begin{equation}
	\hat{\Sigma}_+ =0
	\, ,\quad
	\hat{\sigma}_1 = \frac{1-3\hat{v}_1^2}{3(1-\hat{v}_1^2)}
	\, ,\quad
	\hat{v}_2 = \sqrt{1-\hat{v}_1^2}
	\ .
\end{equation}
The actual expressions for the constants are not important. The free
constants are $\hat{M}$, $\hat{v}_1$, $C_{(1-V^2)}$, $C_{\Sigma_+2}$ and
$C_{\Sigma_+3}$.

The angular variables are given asymptotically by:
\begin{align}
\psi &\approx \hat{\psi}+\frac{\sqrt{3}}{\hat{M}}e^{\tau}, \\
\theta &\approx  \hat{\theta}+ \frac{\sqrt{3}\hat{v}_1}{\hat{M}}e^{\tau}.
\end{align}

The Weyl scalars diverge as
\begin{equation}
        (\Wm,\Wc) \approx - \frac{12}{\hat{M}^2}
	\sqrt{\frac{1-9\hat{v}_1^2}{9(1-\hat{v}_1^2)}} \, e^{2\tau}
        (\cos 2\psi_2,\sin2\psi_2)\ ,
\end{equation}
where $ 0 \leq \hat{v}_1 \leq \frac13$.

\section{Discussion} 
\begin{table}
\centering
\begin{tabular}{|c|c|c|l|}
\hline
 Invariant &   &  &  \\
 subspace & Matter & Attractor & Comments \\ \hline \hline
 $T(VII_0)$ & $2/3<\gamma<4/3$ & $\widetilde{P}_1$ & non-tilted \\
            & $\gamma=4/3$     & $\widetilde{P}_3$ & vortic \\
	    & $4/3<\gamma<2$ & $\widetilde{P}_4$ & $V\rightarrow 1$, vortic \\
	    \hline
$F(VII_0)$ & $2/3<\gamma<4/3$ & $\widetilde{P}_1$ & non-tilted \\
            & $\gamma=4/3$     & $\widetilde{P}_3\big{|}_{v_1=0}$ & vortic \\
	    & $4/3<\gamma<2$ & $\widetilde{P}_4\big{|}_{v_1=0}$ & $V\rightarrow 1$, vortic \\
	    \hline
$T_1(VII_0)$ & $2/3<\gamma\leq 4/3$ & $\widetilde{P}_1$ & non-tilted \\
	    & $4/3<\gamma<2$ & $\widetilde{P}_2$ &  tilted \\
\hline
$B(VII_0)$ & $2/3<\gamma\leq 4/3$ & $\widetilde{P}_1$ &  \\
	    & $4/3<\gamma<2$ & $\widetilde{P}_2\big{|}_{v_1=0}$ &  \\
\hline
\end{tabular}
\caption{The late-time behaviour of the VII$_0$ Bianchi  model with a tilted
$\gamma$-law perfect fluid (see the text for details and
references). The comments refer to the late-time asymptotics, and for all cases 
$\bar{N}\rightarrow \infty$. The case $0<\gamma<2/3$ is covered by the no-hair theorem (the non-tilted version is given in \cite{Wald}, the tilted version in \cite{CH}). 
The value $\gamma=2/3$ is a bifurcation value for which $\bar{N}\rightarrow $ constant. } \label{tab:outline}
\end{table}

We have analysed the asymptotic dynamical behaviour of the tilted Bianchi type
VII$_0$ model at late times. The analytical results are supported by
numerical calculations (which are discussed in Appendix C).
A summary of the late time asymptotic behaviour of the tilted Bianchi type
VII$_0$ model is given in Table \ref{tab:outline}.  The attractors all
refer to the reduced system, which is valid in the limit
$M=1/\bar{N}\rightarrow 0$.
A striking feature of the type VII$_0$ model is the behaviour of the Weyl tensor. 
For $1<\gamma$, the expansion-normalised Weyl scalars $\mathcal{W}_1$ and 
$\mathcal{W}_2$ are unbounded into the future. In the terminology of \cite{BHWeyl},
the model is said to be \emph{extremely Weyl dominant} at late times. 
Note that the (non-expansion-normalised) Weyl invariants actually decay at late 
times; e.g., for $\gamma\geq 4/3$ 
\beq
\left(C_{abcd}C^{abcd},~C_{abcd}{}^*C^{abcd}\right)=48H^4(\Wm,~\Wc) \sim e^{-6\tau}(\cos 2\psi_2,~\sin 2\psi_2).\nonumber 
\eeq

The general tilted type VIII model, which shares many of 
the features of the Bianchi type VII$_0$ model with a tilted fluid studied here,
is currently under investigation. In order to complete the analysis of the
late-time behaviour of  tilted Bianchi models, 
it then remains to study the  Bianchi type VI$_h$ models\footnote{We should stress that some results regarding the stability of certain solutions are already known \cite{BHtilted,CH,Apo}.}.

We have focused on the late-time behaviour of the tilted Bianchi type
VII$_0$ model. However, since the Bianchi type VII$_0$ model
has type II as part of its boundary, it is plausible that
the tilted type VII$_0$ model is chaotic \cite{BKL} at early times
close to the initial singularity.

In this paper we have utilized a formalism adapted to the
timelike geodesics orthogonal to the hypersurfaces defined by the type
VII$_0$ group action.  We note that if we consider the integral
\beq
\Delta T\equiv \int_{\tau_0}^{\infty}\frac 1H\sqrt{1-V^2}{\d \tau}\sim \int_{\tau_0}^{\infty}e^{\frac{8-5\gamma}{2-\gamma}\tau}\d \tau,
\eeq
which corresponds to the proper time of an observer from $\tau=\tau_0$ to $\tau=\infty$ following the fluid congurence, there is a change of behaviour at $\gamma=8/5$.
For models with  $\gamma<8/5$, we find that this integral diverges. However, for
fluids with $\gamma>8/5$, we find that this integral is finite.
Since asymptotically at late times a fluid
with $\gamma>8/5$ will be extremely tilted, this means that the fluid will reach future null infinity in finite proper time. So in 
spite of the fact that these spacetimes are future geodesically complete
\cite{Rendall}, the world-lines defined by the fluid congruences seem to
approach the boundary so quickly that they reach null infinity within finite
proper time as measured by the fluid.  A similiar phenomenon occurs for the
LRS type V model with $\gamma>4/3$ \cite{CollinsEllis}.  The physical 
interpretation of these models is extremely 
complicated.  To fully
understand this behaviour we need to study the dynamics
using a formulation adapted to the fluid (i.e., a fluid-comoving viewpoint).
We shall return to this in future work.

\section*{Acknowledgment}
This work was supported by a Killam Postdoctoral Fellowship (SH) and 
the Natural Sciences and Engineering Research Council of Canada (RJvdH, WCL and AAC). 
\appendix

\section{New variables}
\label{App:A}
Let us introduce the following variables
\beq
{\mbold\Sigma}_{\times}+i{\bf N}_{\times} &\equiv & e^{2i\psi}{\bf X}, \\
{\mbold\Sigma}_{\times}-i{\bf N}_{\times} &\equiv & e^{-2i\psi}{\bf Y}, \\
{\bf v} &\equiv & e^{-i\theta}v_2,
\eeq
where 
\beq
\psi' &=& \sqrt{3}\bar{N},\label{Newpsi}\\
\theta' &=& \sqrt{3}\left\{\bar{N}v_1+\frac 12\mathrm{Im}\left[(1+v_1){\bf X}e^{2i(\theta+\psi)}+ (1-v_1){\bf Y}e^{2i(\theta-\psi)}\right]\right\}.
\eeq
From the remaining freedom we have in choosing the variables and the gauge function we can, 
for example, choose the \emph{initial} values for ${\bf X}$ and ${\bf Y}$ to both be real. 
In any case, objects like $|{\bf X}|^2$ and $|{\bf Y}|^2$ are gauge-independent and are consequently
independent of any such choice.  We also define $M\equiv 1/\bar{N}$. 
\beq
\Sigma_+'&=&
(Q-2)\Sigma_++{3}|{\mbold\Sigma}_1|^2-\frac 12\left(|{\bf X}|^2+|{\bf Y}|^2\right)
+\frac{\gamma\Omega}{2G_+}\left(-2v_1^2+v_2^2\right) \nonumber \\
&& +(\Sigma_++1)\mathrm{Re}\left({\bf X}^*{\bf Y}e^{-4i\psi}\right),\label{NewSigma+} \\
{\bf X}' &=& (Q+\Sigma_+-1){\bf X}+\sqrt{3}{\mbold\Sigma}_1^2e^{-2i\psi}\nonumber \\
&& +\mathrm{Re}\left({\bf X}^*{\bf Y}e^{-4i\psi}\right){\bf X}-(1+\Sigma_+){\bf Y}e^{-4i\psi}+\frac{\sqrt{3}\gamma\Omega v_2^2}{2G_+}e^{-2i(\theta+\psi)},\\
{\bf Y}' &=& (Q+\Sigma_+-1){\bf Y}+\sqrt{3}{\mbold\Sigma}_1^2e^{2i\psi}\nonumber \\
&& +\mathrm{Re}\left({\bf X}^*{\bf Y}e^{-4i\psi}\right){\bf Y}-(1+\Sigma_+){\bf X}e^{4i\psi}+\frac{\sqrt{3}\gamma\Omega v_2^2}{2G_+}e^{-2i(\theta-\psi)}, \\
M'&=&
-M\left[Q+2\Sigma_+
+\mathrm{Re}\left({\bf X}^*{\bf Y}e^{-4i\psi}\right)+\sqrt{3}M\mathrm{Im}\left({\bf Y}^*{\bf
X}e^{4i\psi}\right)\right],\\
\Omega'&=& \frac{\Omega}{G_+}\Big\{2Q-(3\gamma-2)
 +\left[2Q(\gamma-1)-(2-\gamma)-\gamma\mathcal{S}\right]V^2\Big\} \nonumber \\
 && +2\mathrm{Re}\left({\bf X}^*{\bf Y}e^{-4i\psi}\right)\Omega \label{NewOmega}
 \\
v_1' &=& \left(T+2\Sigma_+\right)v_1-2\sqrt{3}\mathrm{Re}\left({\mbold\Sigma}_{1}e^{i\theta}\right)v_2 \nonumber \\
&& +\frac{\sqrt{3}}{2}\mathrm{Re}\left({\bf X}e^{2i(\theta+\psi)}-{\bf Y}e^{2i(\theta-\psi)}\right)v_2^2\\
v_2'&=&\left\{T-\Sigma_+-\frac{\sqrt{3}}{2}\mathrm{Re}\left[(1+v_1){\bf X}e^{2i(\theta+\psi)}+(1-v_1){\bf Y}e^{2i(\theta-\psi)}\right]\right\}v_2,\label{Newv2}
\eeq
where
\beq
Q&=& 2(\Sigma^2_++|{\mbold\Sigma}_1|^2)+\frac{1}{2}\left(|{\bf X}|^2+|{\bf Y}|^2\right)+\frac
12\frac{(3\gamma-2)+(2-\gamma)V^2}{1+(\gamma-1)V^2}\Omega.
\eeq
These variables are subject to the constraints
\beq
1&=& \Sigma^2_++\left|{\mbold\Sigma}_1\right|^2+\frac{1}{2}\left(|{\bf X}|^2+|{\bf Y}|^2\right)+\Omega \\
0 &=& \frac{1}{2}\left(|{\bf X}|^2-|{\bf Y}|^2\right)-\frac{\gamma\Omega v_1}{G_+}. \label{NewConst3}
\eeq 
We can also split $V^2{\mathcal{S}}$ and $T$ into oscillatory, and non-oscillatory parts: 
\beq
V^2\mathcal{S}&=& \Sigma_+\left(-2v_1^2+v_2^2\right)\nonumber \\ && 
+\frac{\sqrt{3}}{2}\mathrm{Re}\left[\left({\bf X}e^{2i(\theta+\psi)}+{\bf Y}e^{2i(\theta-\psi)}\right)v_2^2+4{\mbold\Sigma_1}e^{i\theta}v_1v_2\right],\\
T & =& \frac{(3\gamma-4)(1-V^2)+(2-\gamma)\Sigma_+\left(-2v_1^2+v_2^2\right)}{1-(\gamma-1)V^2}\nonumber\\
&&+ \frac{\sqrt{3}(2-\gamma)}{2(1-(\gamma-1)V^2)}\mathrm{Re}\left[\left({\bf X}e^{2i(\theta+\psi)}+{\bf Y}e^{2i(\theta-\psi)}\right)v_2^2+4{\mbold\Sigma_1}e^{i\theta}v_1v_2\right].
\eeq
\subsection{Assuming $M\rightarrow 0$}
Let us assume that $M\rightarrow 0$ in the following. To investigate the asymptotic dynamics 
we choose a new set of variables. If $x$ is a generic variable, we define $\hat{x}$ as:
\beq
\hat{x}=\frac{x}{1+Mf}-Mg,
\label{hatted}\eeq
where $f$ and $g$ are bounded functions of the state space variables.
If the equation of motion for $x$ is
\beq
x'=Fx+G, \quad F,G \text{ bounded},
\eeq 
we obtain 
\beq
\hat{x}'=\frac{(F-Mf')\hat{x}+G-Mg'}{1+Mf}+MB,
\eeq
where $B$ is a bounded function into the future.  The trick now is to choose the 
functions $f$ and $g$ in order to get rid of the oscillatory terms and write the system in the following form: 
\beq
\hat{x}'=\hat{F}\hat{x}+\hat{G}+\hat{M}\hat{B}, 
\label{hattedsystem}\eeq
where the function $\hat{B}$ is bounded to the future. For the variable $\hat{M}$, we choose the following form:
\beq
\hat{M}'=\hat{F}\hat{M}+\hat{M}^2\hat{B}.
\eeq

Using the transformation eq.(\ref{hatted}), the following functions $f$ and $g$ will do the trick for the various variables:
\beq
\Sigma_+: && f=g=-\frac{1}{4\sqrt{3}}\mathrm{Im}\left({\bf X}^*{\bf Y}e^{-4i\psi}\right), \nonumber \\
{\bf X}: && f=-\frac{1}{4\sqrt{3}}\mathrm{Im}\left({\bf X}^*{\bf Y}e^{-4i\psi}\right), \nonumber \\
&& g=-\frac{i}{4\sqrt{3}}(1+\Sigma_+){\bf Y}e^{-4i\psi}+\frac{i\gamma\Omega v_2^2}{4G_+(1+v_1)}e^{-2i(\theta+\psi)},\nonumber \\
{\bf Y}: && f=-\frac{1}{4\sqrt{3}}\mathrm{Im}\left({\bf X}^*{\bf Y}e^{-4i\psi}\right), \nonumber \\
&& g=\frac{i}{4\sqrt{3}}(1+\Sigma_+){\bf X}e^{4i\psi}-\frac{i\gamma\Omega v_2^2}{4G_+(1-v_1)}e^{-2i(\theta-\psi)},\nonumber \\
M: && f=\frac{1}{4\sqrt{3}}\mathrm{Im}\left({\bf X}^*{\bf Y}e^{-4i\psi}\right), \quad g=0, \nonumber \\
\Omega: && f=-\frac{1}{2\sqrt{3}}\mathrm{Im}\left({\bf X}^*{\bf Y}e^{-4i\psi}\right)-\frac{\gamma }{4G_+}\mathrm{Im}\left[\frac{{\bf X}v_2^2}{1+v_1}e^{2i(\theta+\psi)}-\frac{{\bf Y}v_2^2}{1-v_1}e^{2i(\theta-\psi)}\right],\nonumber \\ &&g=0,\nonumber \\
v_1:&& f=\frac{(2-\gamma)}{4G_-}\mathrm{Im}\left[\frac{{\bf X}v_2^2}{1+v_1}e^{2i(\theta+\psi)}-\frac{{\bf Y}v_2^2}{1-v_1}e^{2i(\theta-\psi)}\right],\nonumber \\
&& g=-\frac{1}{4}\mathrm{Im}\left[\frac{{\bf X}v_2^2}{1+v_1}e^{2i(\theta+\psi)}+\frac{{\bf Y}v_2^2}{1-v_1}e^{2i(\theta-\psi)}\right], \nonumber \\
v_2: && f=\frac{(2-\gamma)}{4G_-}\mathrm{Im}\left[\frac{{\bf X}v_2^2}{1+v_1}e^{2i(\theta+\psi)}-\frac{{\bf Y}v_2^2}{1-v_1}e^{2i(\theta-\psi)}\right]-\frac{1}{4}\mathrm{Im}\left[{\bf X}e^{2i(\theta+\psi)}-{\bf Y}e^{2i(\theta-\psi)}\right], \nonumber \\
&& g=0.\nonumber
\eeq
We note that all of the above functions $f$ and $g$ are bounded; in particular, 
\[ 0\leq \frac{v_2^2}{1\pm v_1}=\frac{V^2-v_1^2}{1-v_1^2}(1\mp v_1)\leq 2.\]

Let $h^i$ be the transformation $h^i:~x^i\mapsto \hat{x}^i$ given above 
(supplemented with $(\hat\theta,\hat\psi)=(\theta,\psi)$). We note that if there exists an $\epsilon>0$ such that $|v_1|<1-\epsilon$, then the Jacobian ${\bf J}\equiv \left(\frac{\partial h^i}{\partial x^j}\right)$ has determinant
\beq
\det({\bf J})=1+\sum_{n=1}^kM^nb_n, \quad b_n \text{ bounded.}
\eeq
Hence, by the inverse function theorem, if ${M}\rightarrow 0$ there will exist a $t_0$ such that
$h^i:~x^i\mapsto \hat{x}^i$ has a continuous inverse for all $t>t_0$.  This means that close to
the equilibrium points of the reduced system, the functions $h^i$ and their inverse are well
defined.  We can consequently rewrite the system as eq.(\ref{hattedsystem}) close to the reduced
system.  As $M\rightarrow 0$, we can then use centre manifold theory to show that the
system effectively reduces to the reduced system of section \ref{sect:redsys}.

\section{Bounds on oscillatory terms}
\label{App:B}
Consider an equilibrium point, ${\sf X}_0$, for the reduced system. If we linearise the 
full system about ${\sf X}_0$ (treating $\theta$ and $\psi$ as sources and not part of the system) we get a matrix equation of the form
\beq
{\sf X}'=\left({\bf L}+{\bf S}(\theta,\psi)\right){\sf X}+{\sf P}(\theta,\psi).
\eeq
Here, the matrix ${\bf S}(\theta,\psi)$ and the vector ${\sf P}(\theta,\psi)$ contains the 
oscillating terms that need to be estimated. 

This linearised equation has the formal solution 
\beq
{\sf X}(t_2)=e^{{\bf I}(t_1,t_2)}{\sf X}(t_1)+e^{{\bf
    I}(t_1,t_2)}\int_{t_1}^{t_2}e^{-{\bf I}(t_1,t)}{\sf P}(\theta,\psi)\d t, 
\eeq
where 
\[ {\bf I}(t_1,t_2)=\int_{t_1}^{t_2}\left({\bf L}+{\bf S}(\theta,\psi)\right)\d t.\] 

In order to estimate the contribution from the oscillating terms we need the following result: 
\begin{lem}
Consider the integral 
\beq
I(t_1,t_2)\equiv \int_{t_1}^{t_2}e^{\lambda t}e^{i\beta e^{\rho t}}\d t, \quad \beta,\rho >0, \quad \rho>\lambda.
\eeq
Then there exists a constant $D$ such that for every $t_1$ and $t_2$ satisfying 
\[ t_1<t_2, \]
we have 
\beq 
\left|I(t_1,t_2)\right|\leq D\left(e^{(\lambda-\rho)t_1}+e^{(\lambda-\rho)t_2}\right).
\eeq
\end{lem} 
\begin{proof}
We introduce the variable $z=e^{\rho t}$ so that the integral can be written
\[ I=\frac{1}{\rho}\int_{z_1}^{z_2}z^{\frac{\lambda-\rho}{\rho}}e^{i\beta z}\d z.\] 
Considering $z$ as a complex variable, we notice that the integrand only has (at most) a 
singularity at $z=0$. Hence, the integrand will be analytic in the entire half-plane 
$\mathrm{Re}(z)>0$. By choosing a suitable closed path we can rewrite $I$ as: 
\beq
I=-I_{c_1}-I_{c_2}, \quad c_i=z_i+iy, ~0\leq y<\infty.
\eeq
These two integrals can be estimated:
\beq
\rho \left|I_{c_1}\right|=\left|\int_{0}^{\infty}(z_1+iy)^{\frac{\lambda-\rho}{\rho}}e^{i\beta z_1-\beta y}\d y\right|\leq \int_{0}^{\infty}(z_1^2+y^2)^{\frac{\lambda-\rho}{2\rho}}e^{-\beta y}\d y.
\eeq
Since $\lambda<\rho$ the factor $(z_1^2+y^2)^{\frac{\lambda-\rho}{2\rho}}$ is monotonically decreasing for $y>0$; hence, 
\beq
\rho \left|I_{c_1}\right|\leq z_1^{\frac{\lambda-\rho}{\rho}}\int_0^{\infty}e^{-\beta y}\d y=\frac{1}{\beta}z_1^{\frac{\lambda-\rho}{\rho}}. 
\eeq
Thus we have 
\beq 
\left|I\right|&\leq& \left|I_{c_1}\right|+\left|I_{c_2}\right|\leq \frac{1}{\rho \beta}\left(z_1^{\frac{\lambda-\rho}{\rho}}+z_2^{\frac{\lambda-\rho}{\rho}}\right)
= \frac{1}{\rho \beta}\left(e^{(\lambda-\rho)t_1}+e^{(\lambda-\rho)t_2}\right),
\eeq
which proves the Lemma.
\end{proof}

Consider first the linearised $M$ equation with respect to the future
stable points of the reduced equation. We assume that
$M\rightarrow 0$, and by a similar complex integration we get 
\beq
M(t_2)=e^{-\rho(t_2-t_1)}M(t_1)+\mathcal{O}(M^2), \quad \rho
  =\begin{cases} \frac 12(3\gamma-2), & \frac 23 <\gamma<\frac 43 \\
 1, & \frac 43 \leq \gamma <2.\end{cases}
\eeq
Hence, asymptotically, $M\propto e^{-\rho t}$. 

The matrix ${\bf S}(\theta,\psi)$ contains oscillating terms like
$e^{i\alpha}$, where $\alpha$ is a linear combination of $\theta$ and
$\psi$. Using the equation for $\theta$ and $\psi$, and the leading
order approximation for $M$, we obtain
\beq
{\bf I}(t_1,t_2)=\int_{t_1}^{t_2}({\bf L}+ {\bf S}(\theta,\psi))\d
t={\bf L}(t_2-t_1)+{\mathcal O}(e^{-\rho t_1}+e^{-\rho t_2}).
\eeq  
Let us expand ${\sf X}$ (and {\sf P}) in terms of a sum of eigenvectors ${\sf
  X}_\lambda$ of ${\bf L}$; i.e.
\[ {\sf X}=\sum_{\lambda}{\sf X}_\lambda, \quad {\bf L}{\sf X}_\lambda=\lambda{\sf X}_\lambda.\] 
The solution can now be estimated: 
\beq
{\sf X}_{\lambda}(t_2)=e^{\lambda (t_2-t_1)}{\sf
    X}_{\lambda}(t_1)+\mathcal{O}(e^{-\rho
    t_1+\lambda(t_2-t_1)}+e^{-\rho t_2}).
\eeq
This result shows that the deviation from the solution of the reduced
system is of the order of $M$. Hence, since $M\rightarrow 0$, the
linearised solution will asymptotically approach the solution of the
reduced system. 

\section{Numerical analysis}
\label{app:num}
Numerical calculations can be used to both confirm and clarify the
analytical calculations found in the bulk of the paper.  We have
chosen to use the new variables found in Appendix A, in which the
oscillatory part of the system of differential equations can essentially be
isolated from the non-oscillatory part through the
introduction of the variables $\psi$ and $\theta$. Equations (\ref{Newpsi})--(\ref{Newv2}) 
(minus equation (\ref{NewOmega})) were numerically integrated using the `ODE23t'
ODE solver in Matlab with an Absolute Tolerance of $10^{-6}$ and a
Relative Error of $10^{-6}$. The constraint equation (\ref{NewConst3}) was used
to determine if and when the numerical calculations broke down.

\begin{figure}
\includegraphics[width=5.5cm]{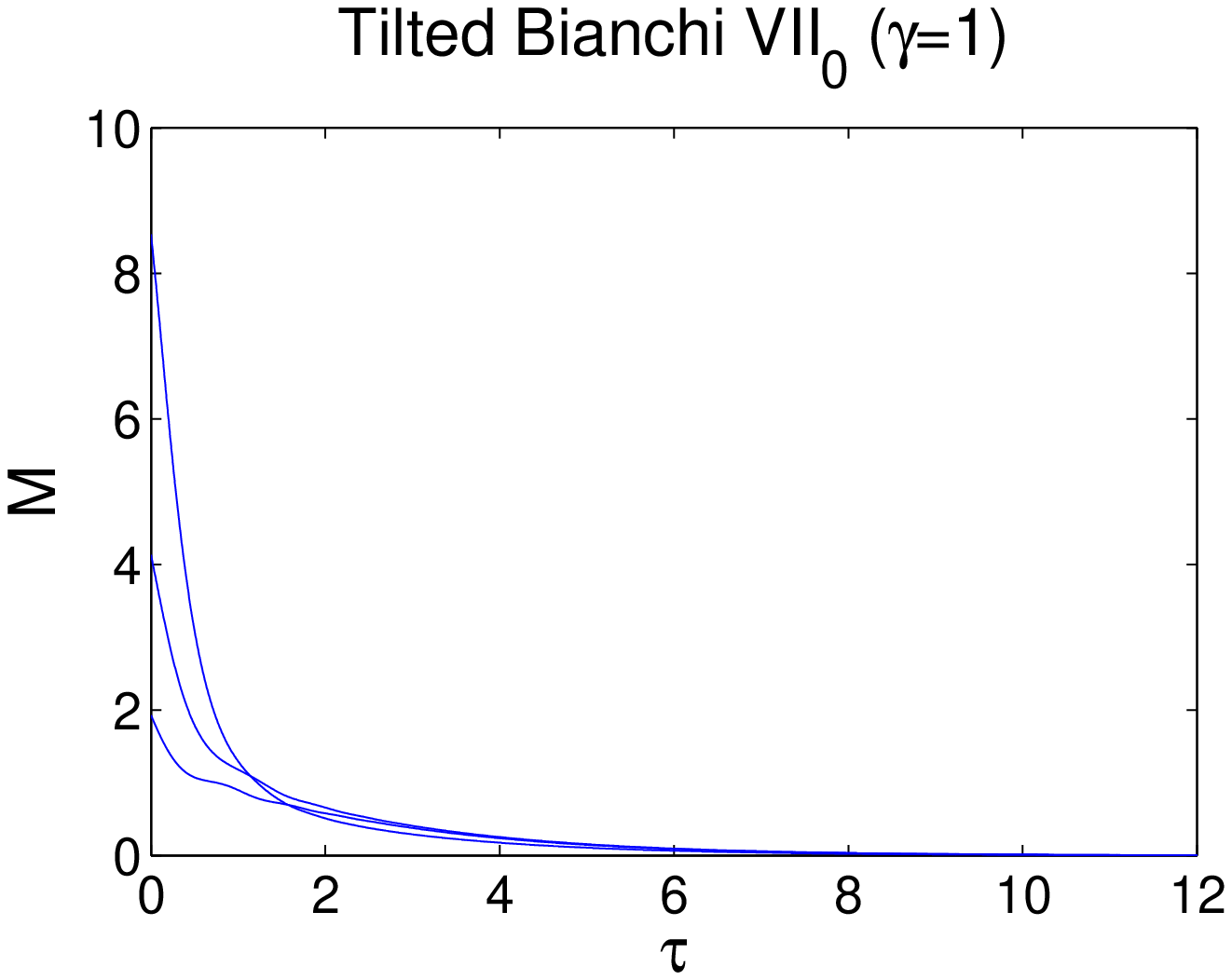}\includegraphics[width=5.5cm]{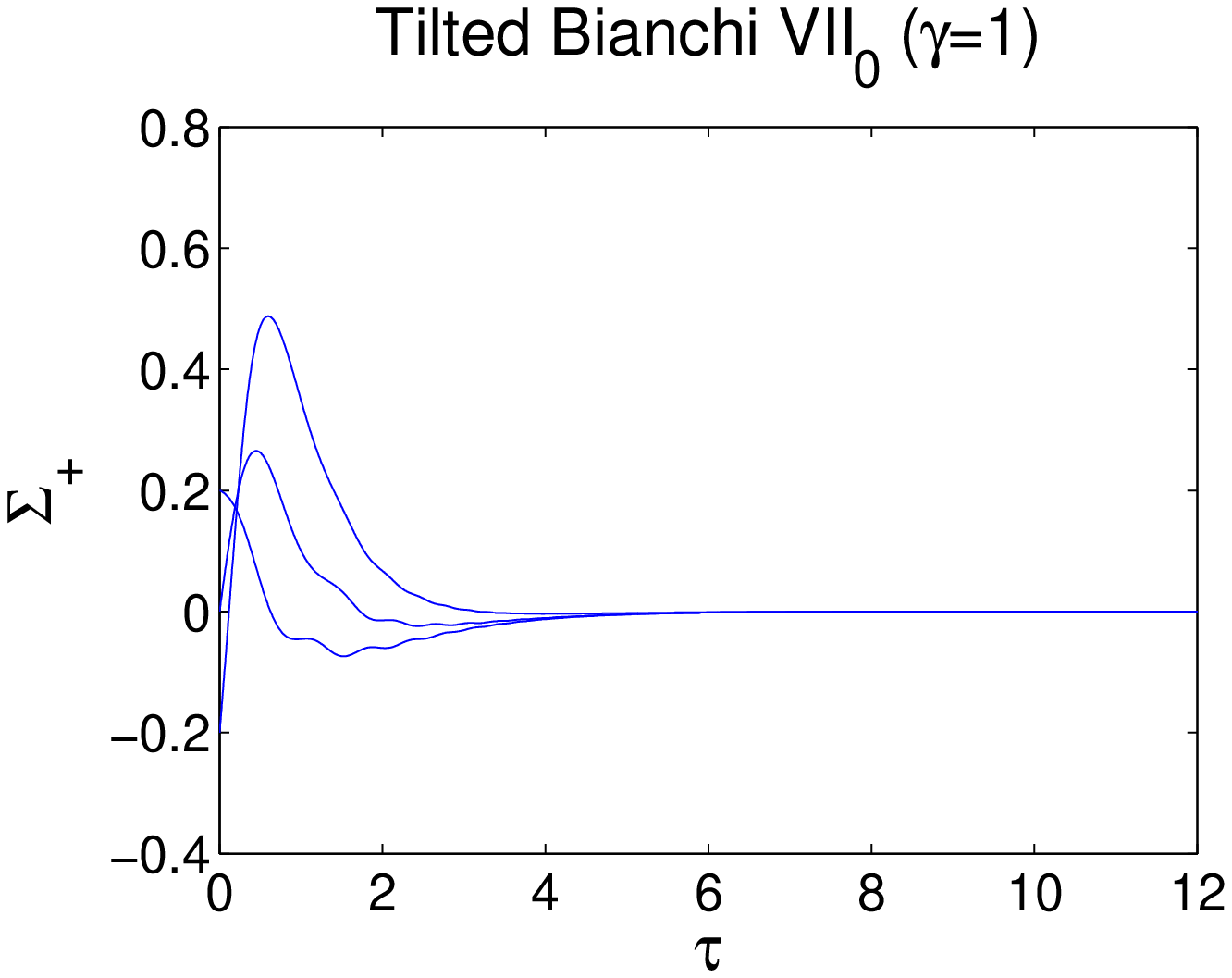}

\includegraphics[width=5.5cm]{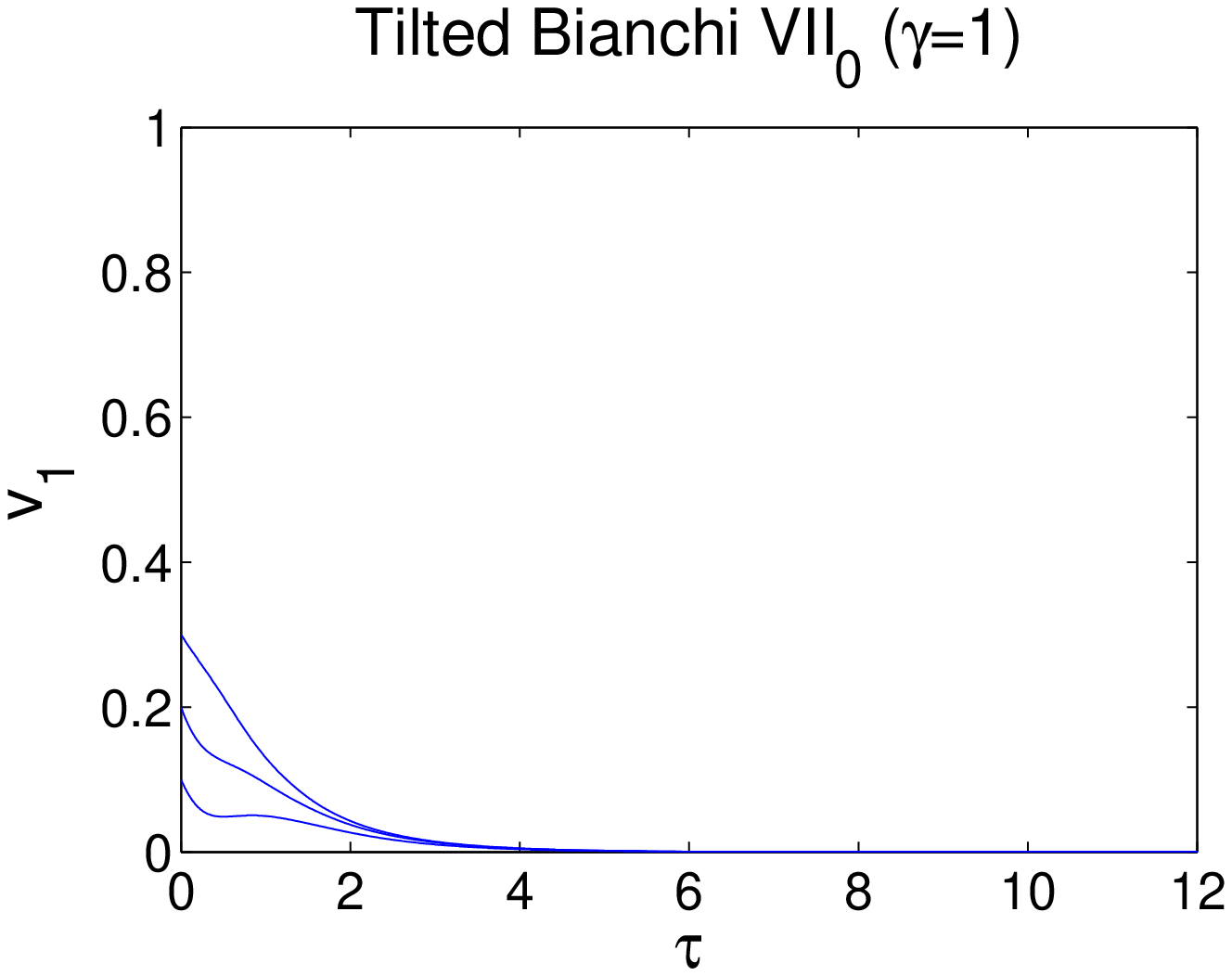}\includegraphics[width=5.5cm]{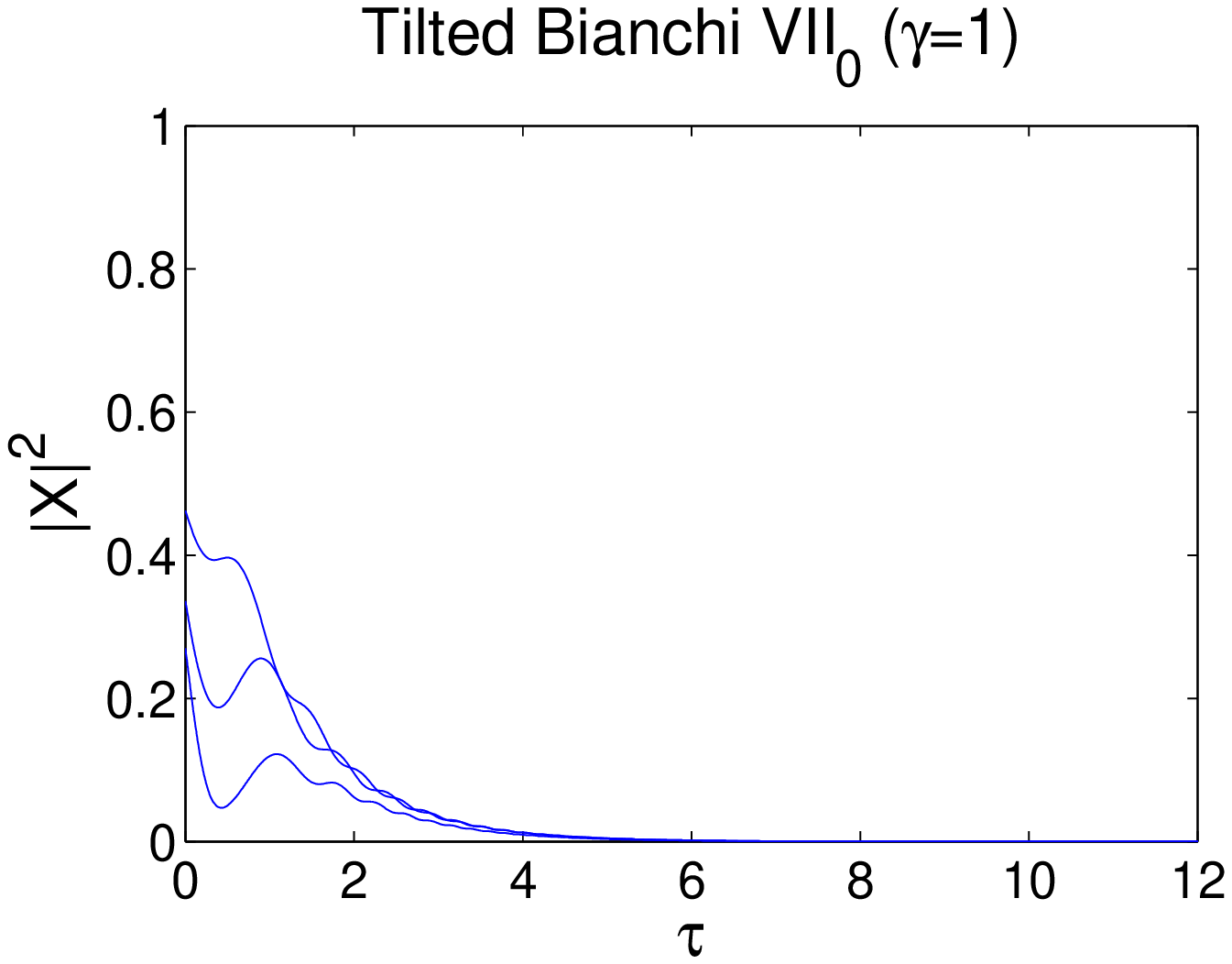}
\caption{Plots of $M$, $\Sigma_+$, $v_1$ and $|{\bf X}|^2$ for
$\gamma=1$, illustrating the rate at which $M$ (and all other
variables) tend to zero.}\label{Fig:g=1}
\end{figure}

\begin{figure}
\includegraphics[width=5.5cm]{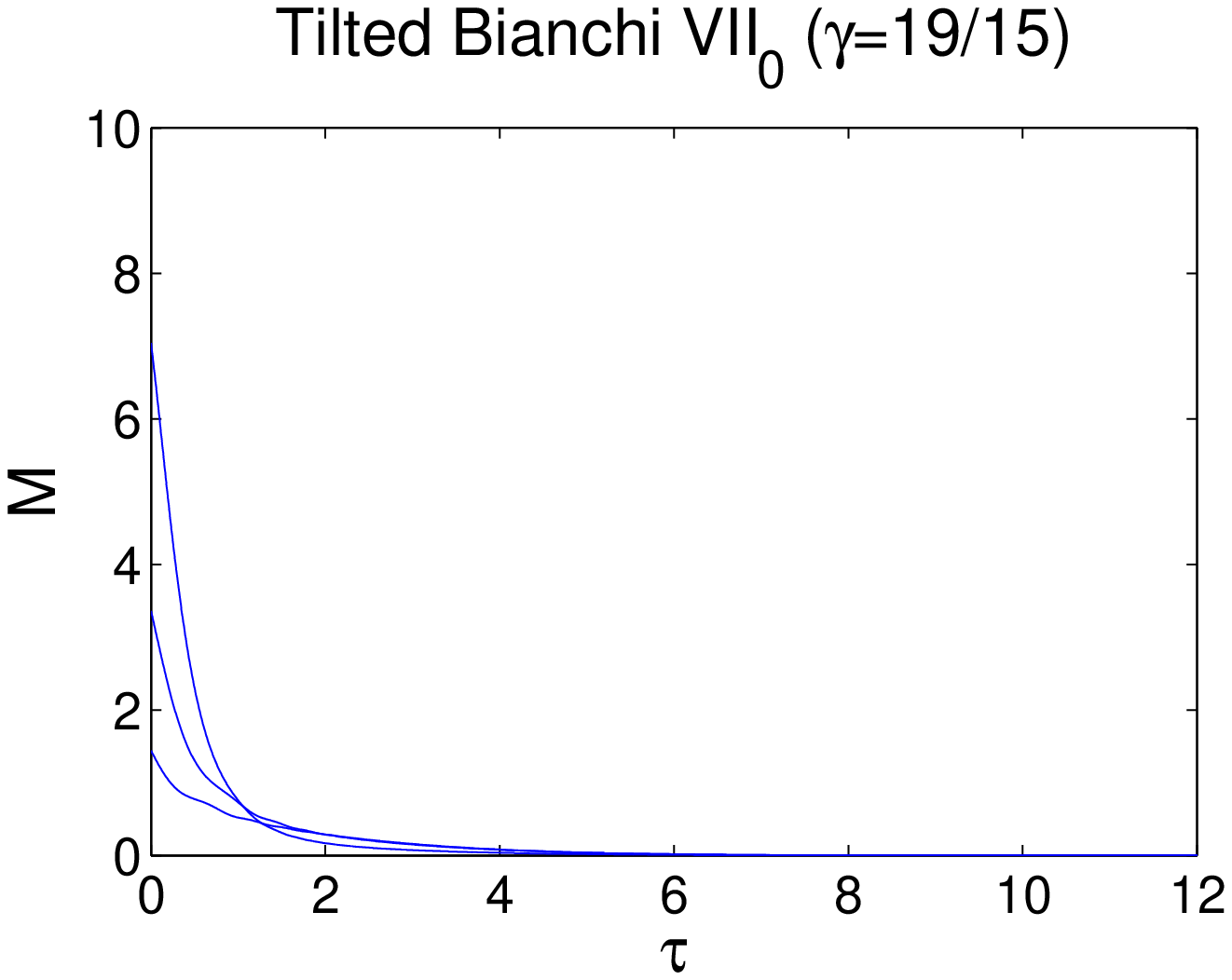}\includegraphics[width=5.5cm]{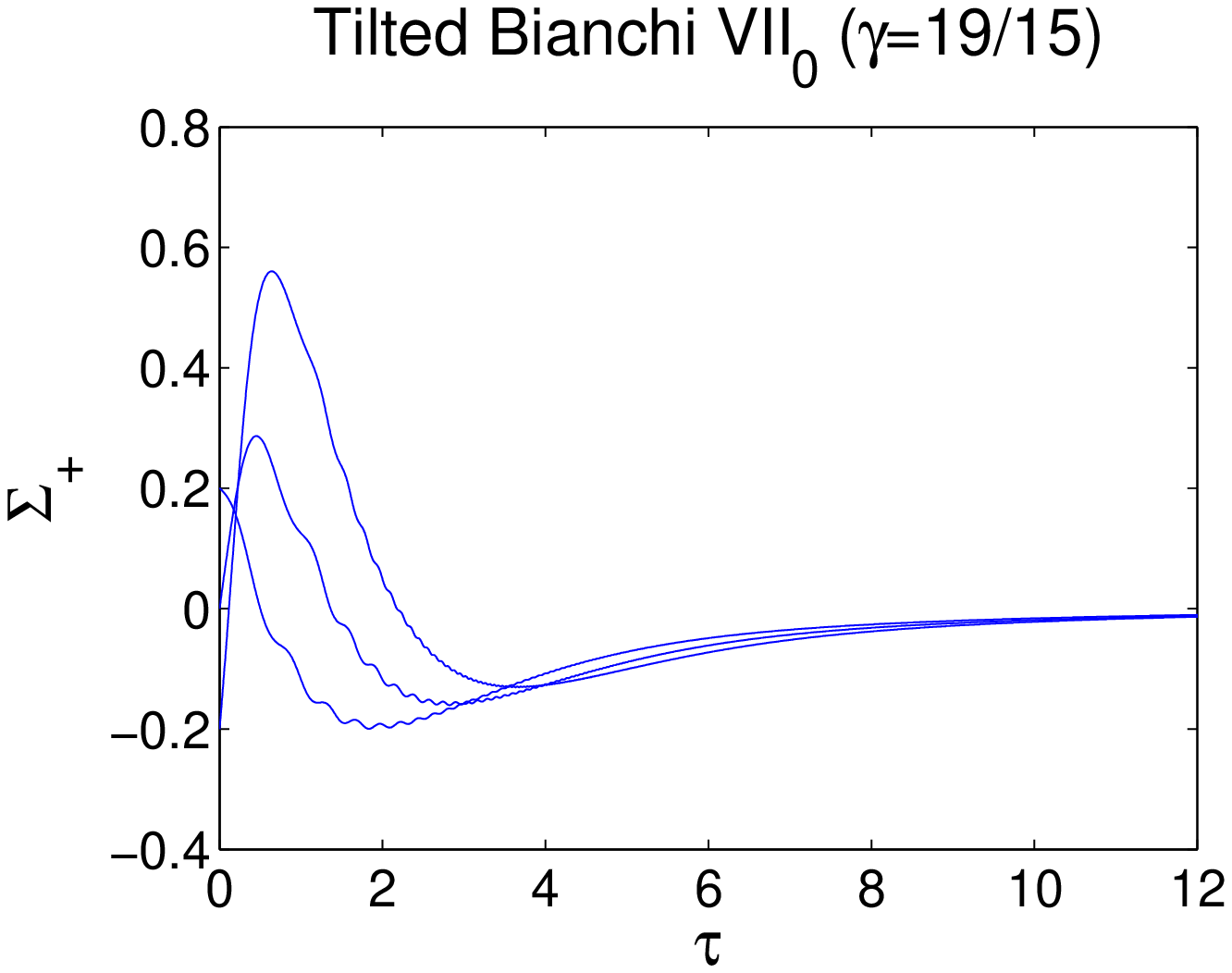}

\includegraphics[width=5.5cm]{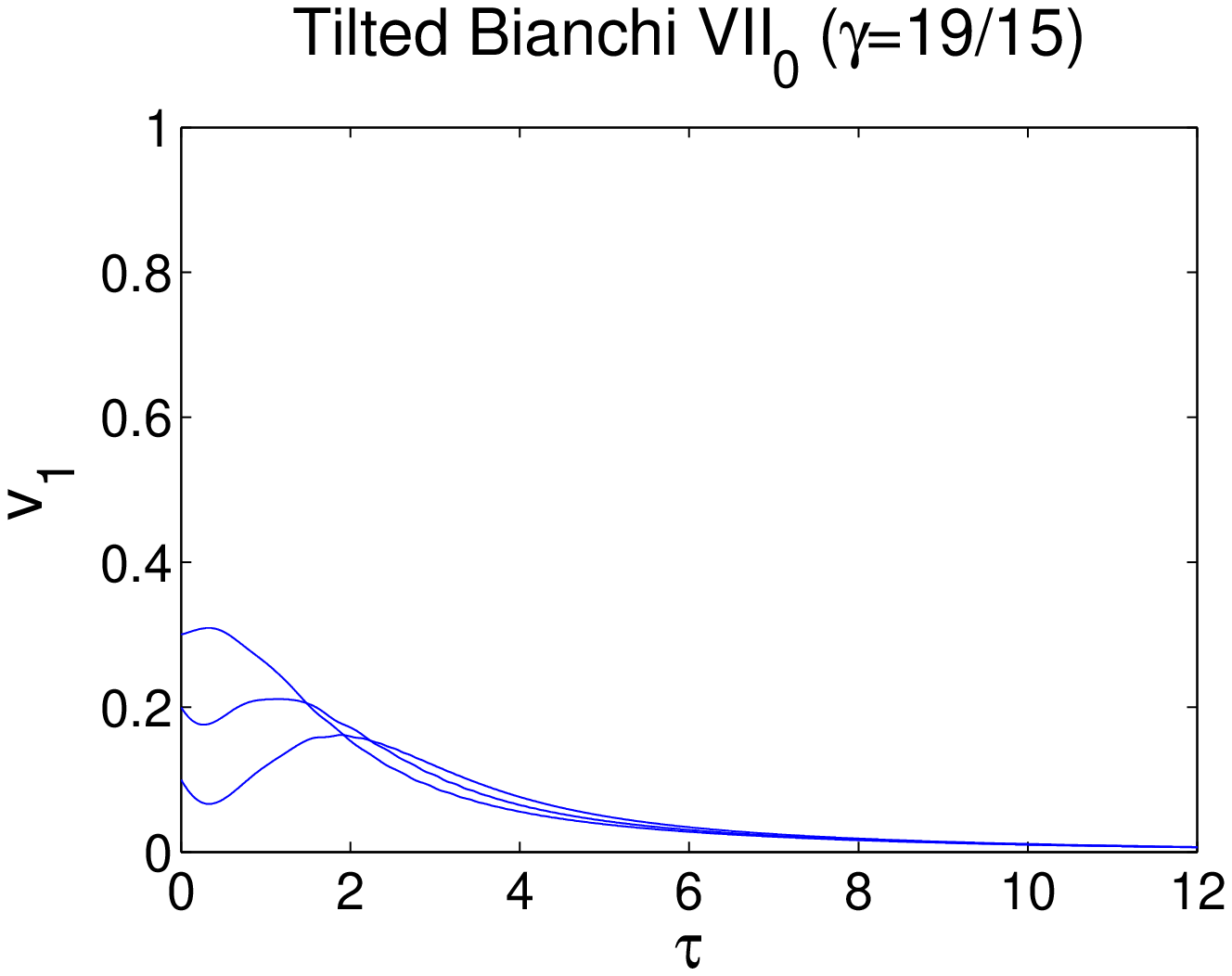}\includegraphics[width=5.5cm]{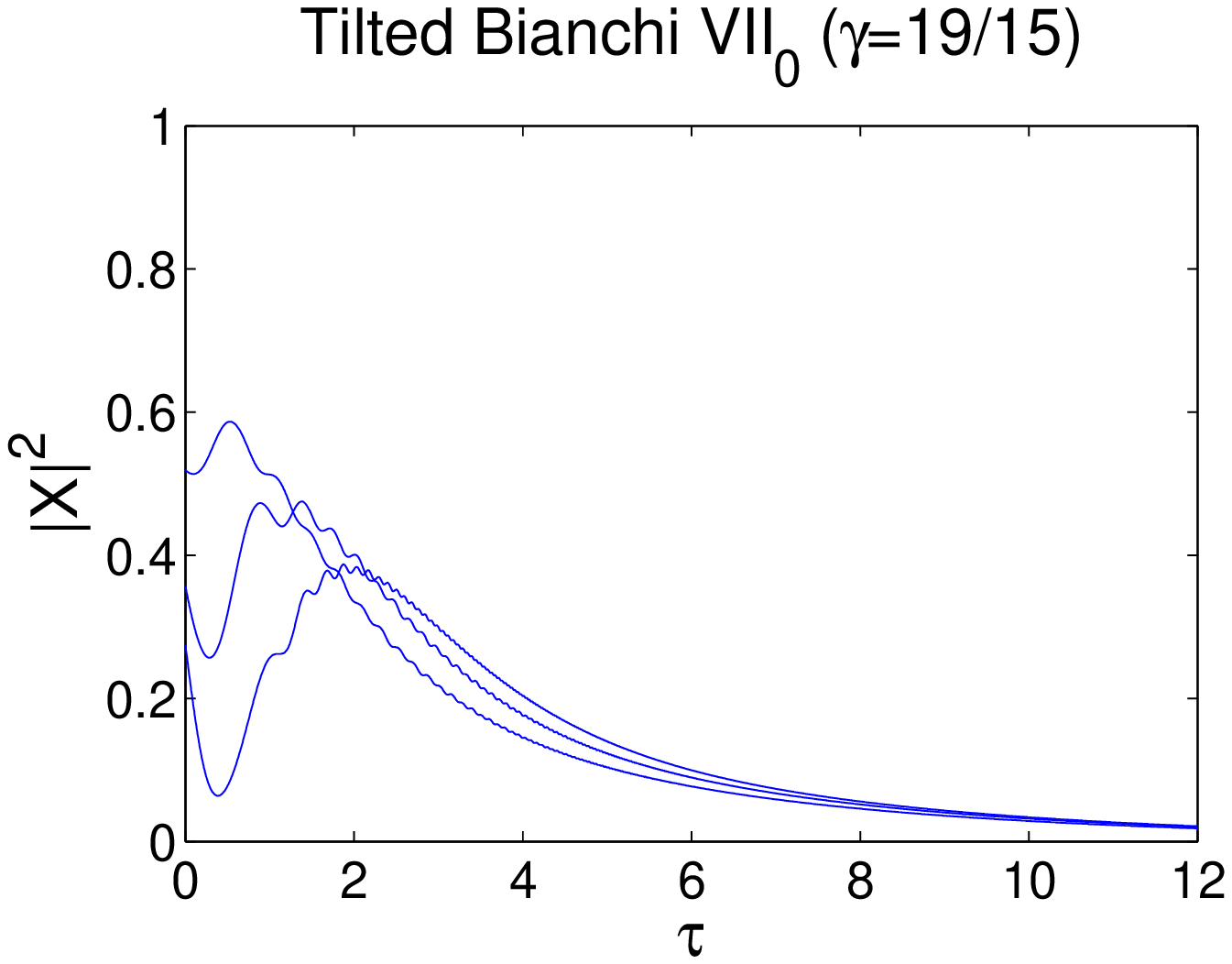}
\caption{Plots of $M$, $\Sigma_+$, $v_1$ and $|{\bf X}|^2$ for
$\gamma=19/15$, illustrating the rate at which $M$ tends to zero.
 Compare the slow approach of $\Sigma_+$,$v_1$ and $|{\bf X}|^2$ towards
 zero with the corresponding plots with $\gamma=1$.} \label{Fig:g=1915}\end{figure}

\begin{figure}
\includegraphics[width=5.5cm]{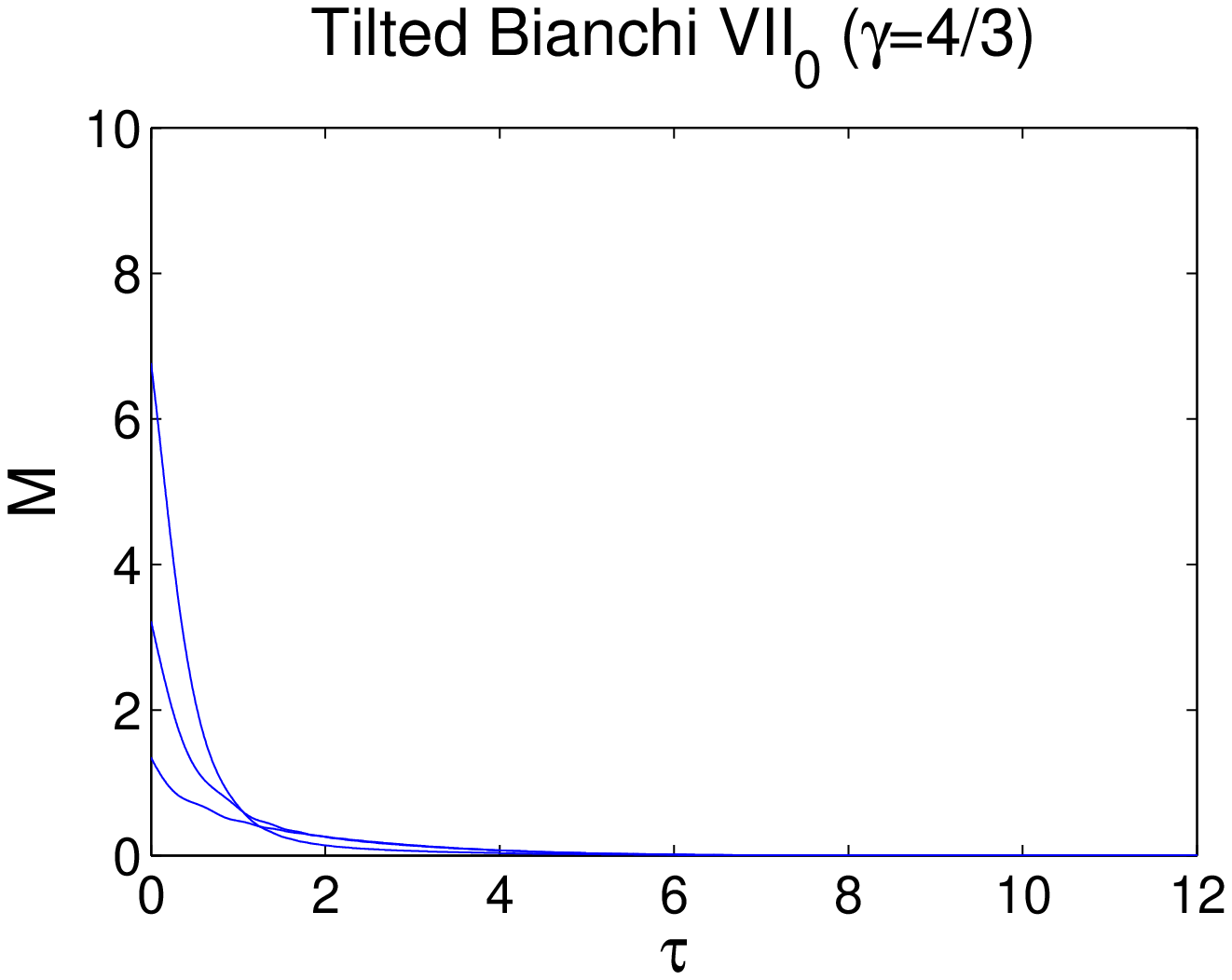}\includegraphics[width=5.5cm]{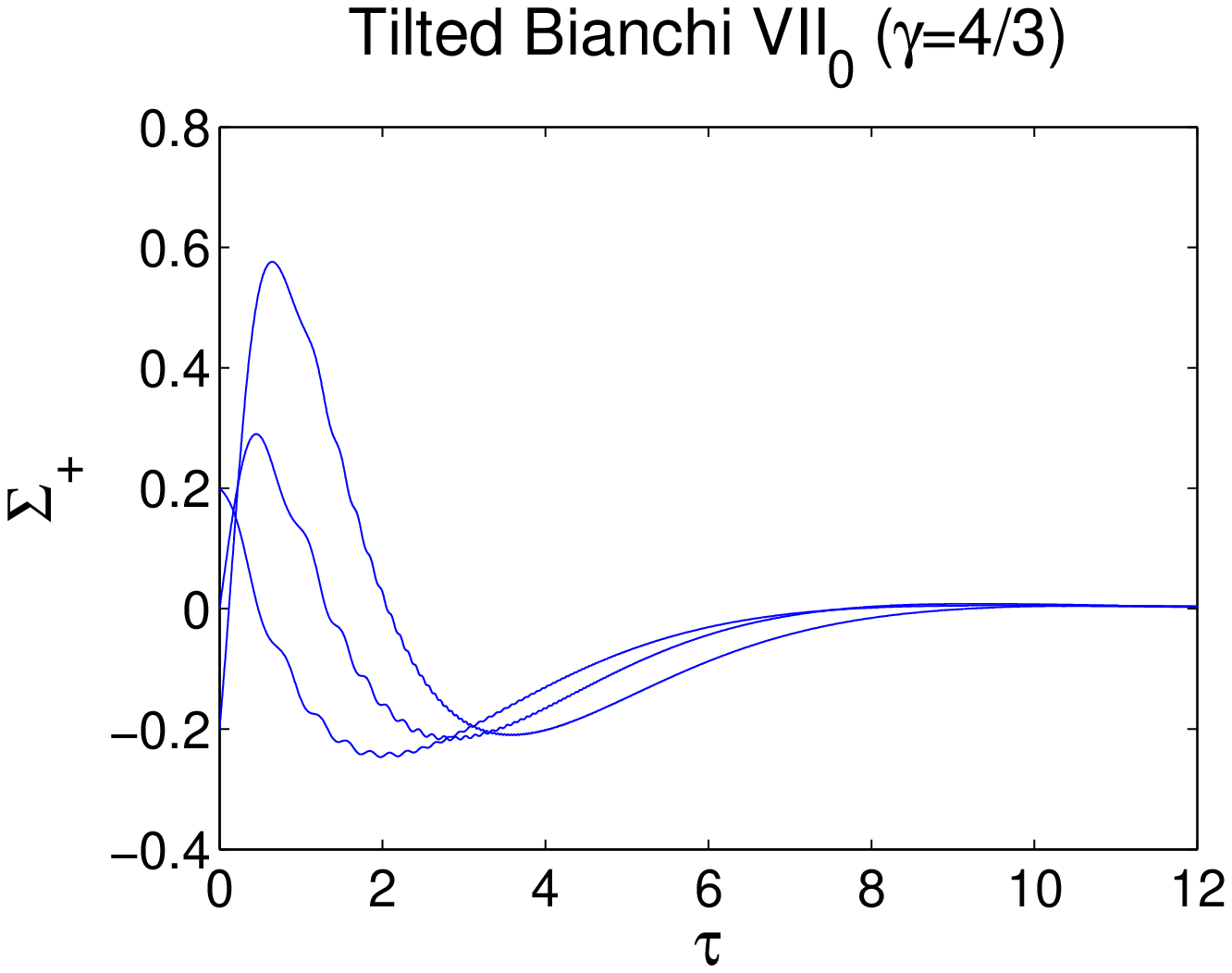}

\includegraphics[width=5.5cm]{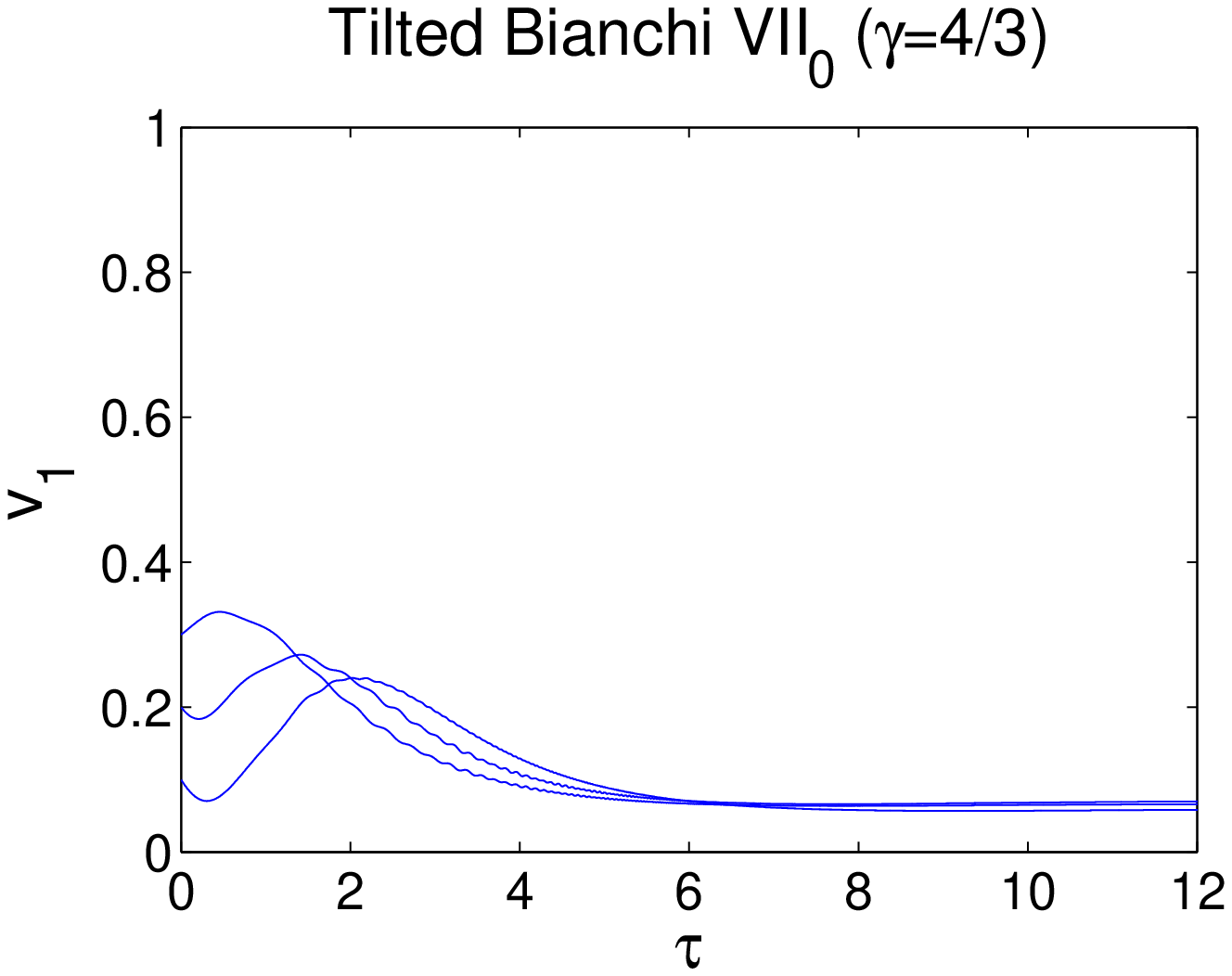}\includegraphics[width=5.5cm]{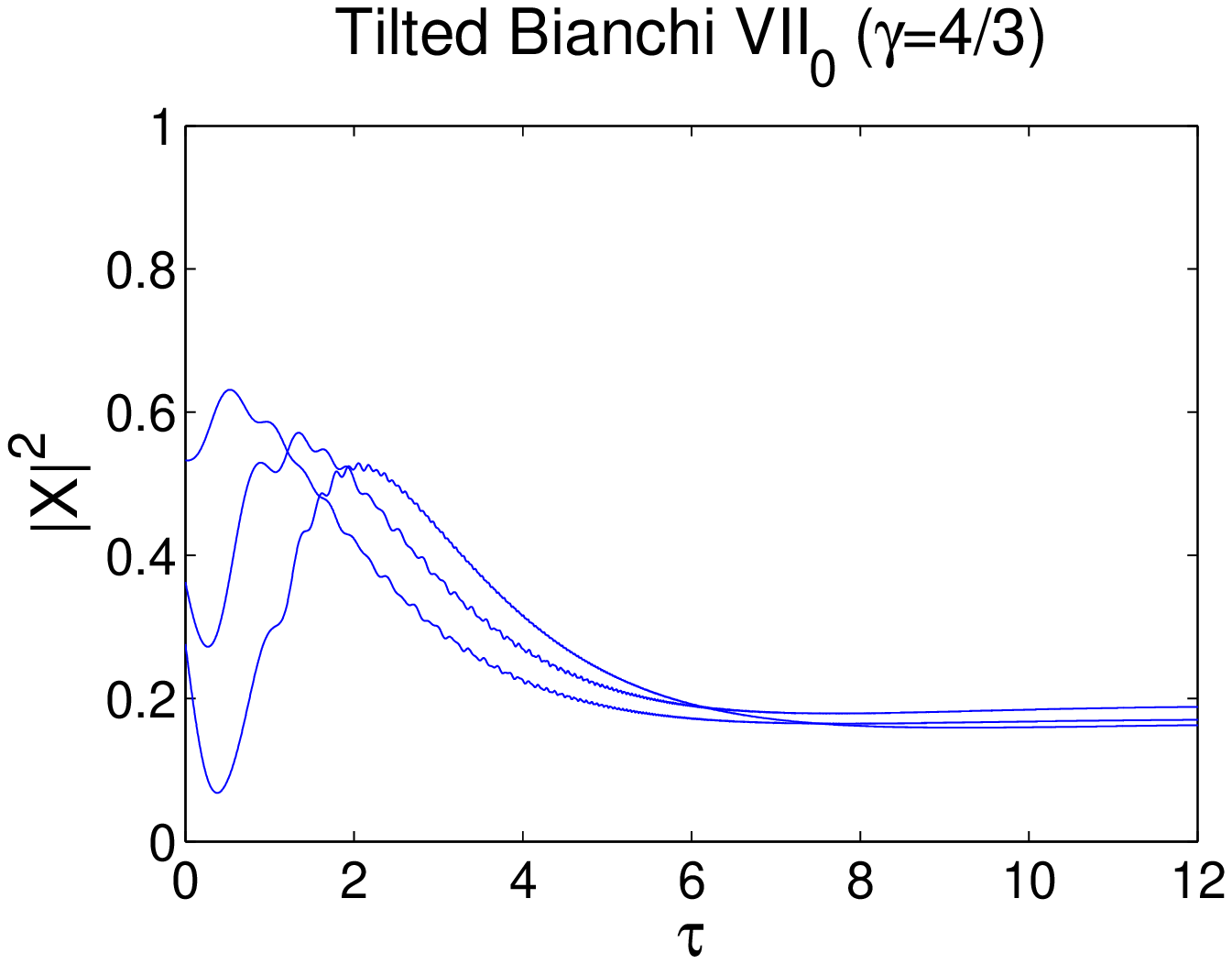}
\caption{Plots of $M$, $\Sigma_+$, $v_1$ and $|{\bf X}|^2$ for
$\gamma=4/3$, illustrating the rate at which $M$ tends to zero.
 Note how the variables $v_1$ and $|{\bf X}|^2$ do
 not approach a unique value.}\label{Fig:g=43}
 \end{figure}

\begin{figure}
\includegraphics[width=5.5cm]{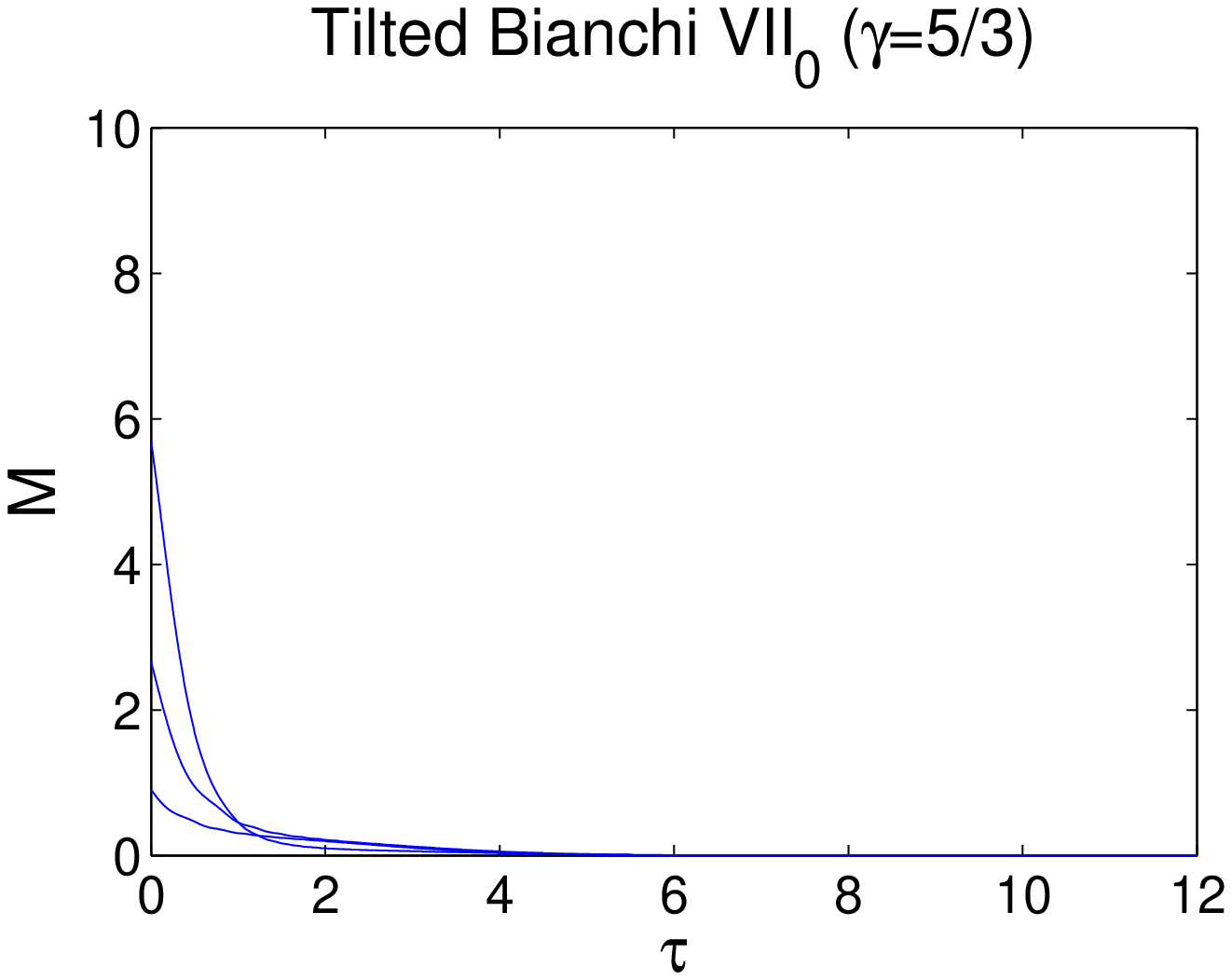}\includegraphics[width=5.5cm]{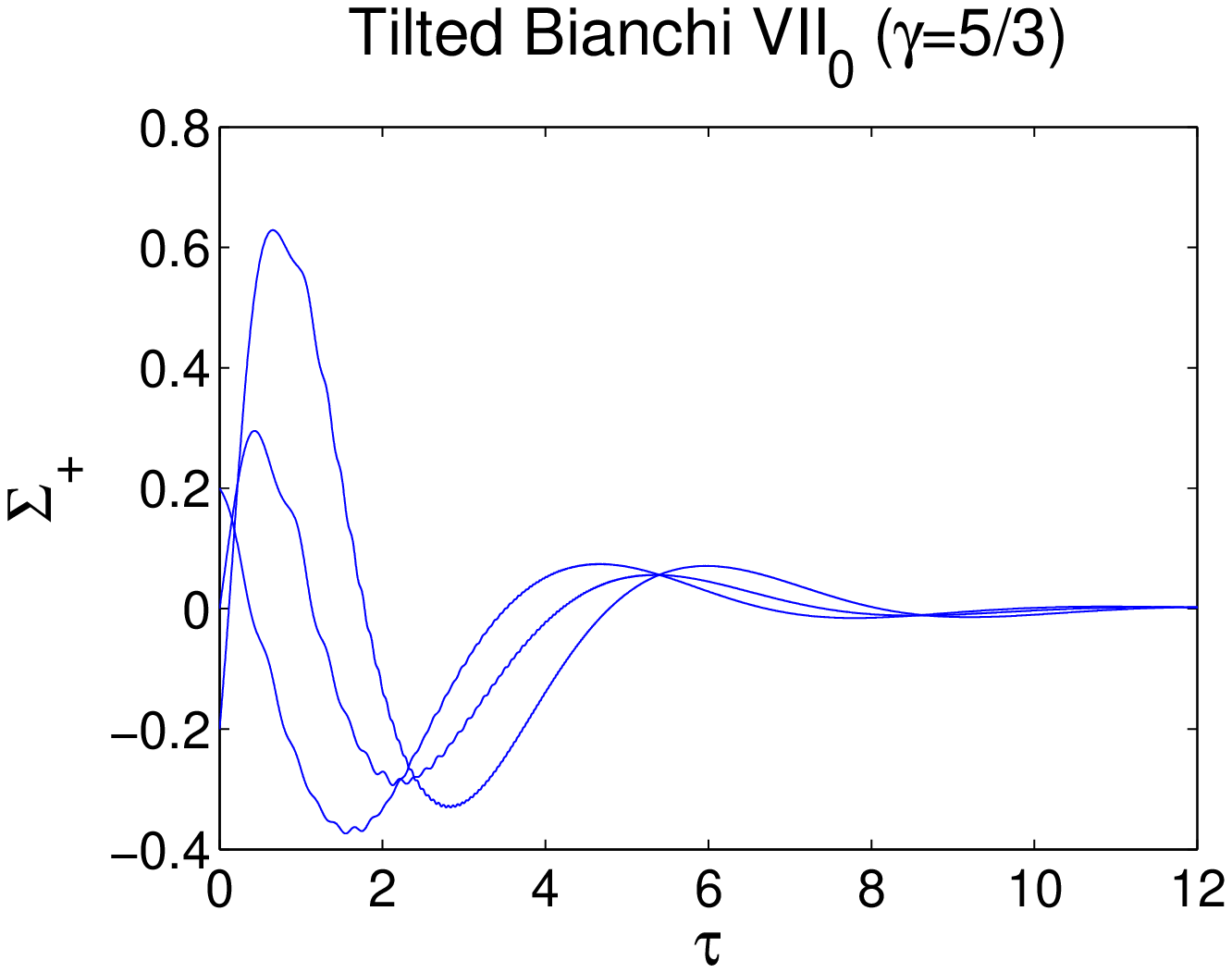}

\includegraphics[width=5.5cm]{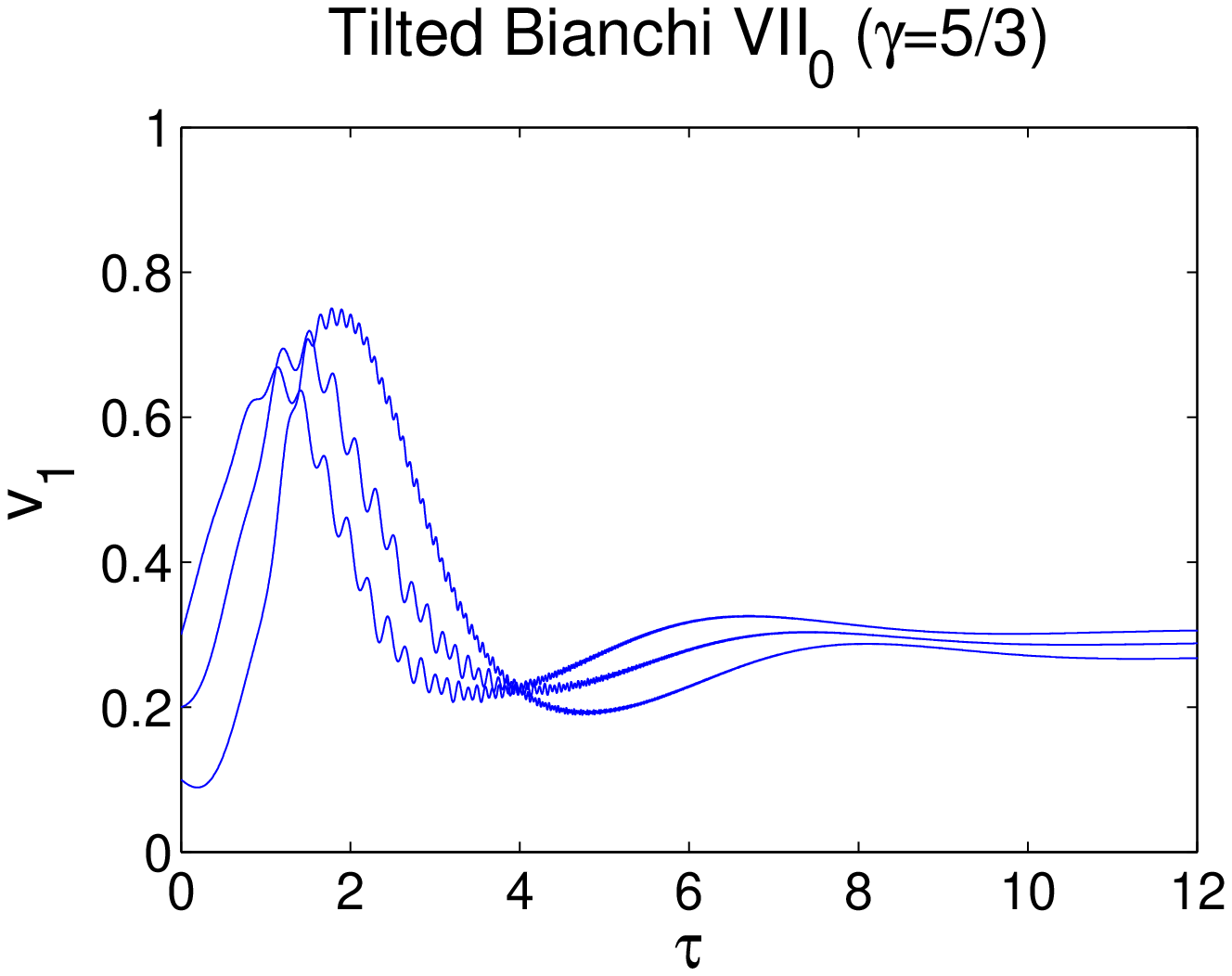}\includegraphics[width=5.5cm]{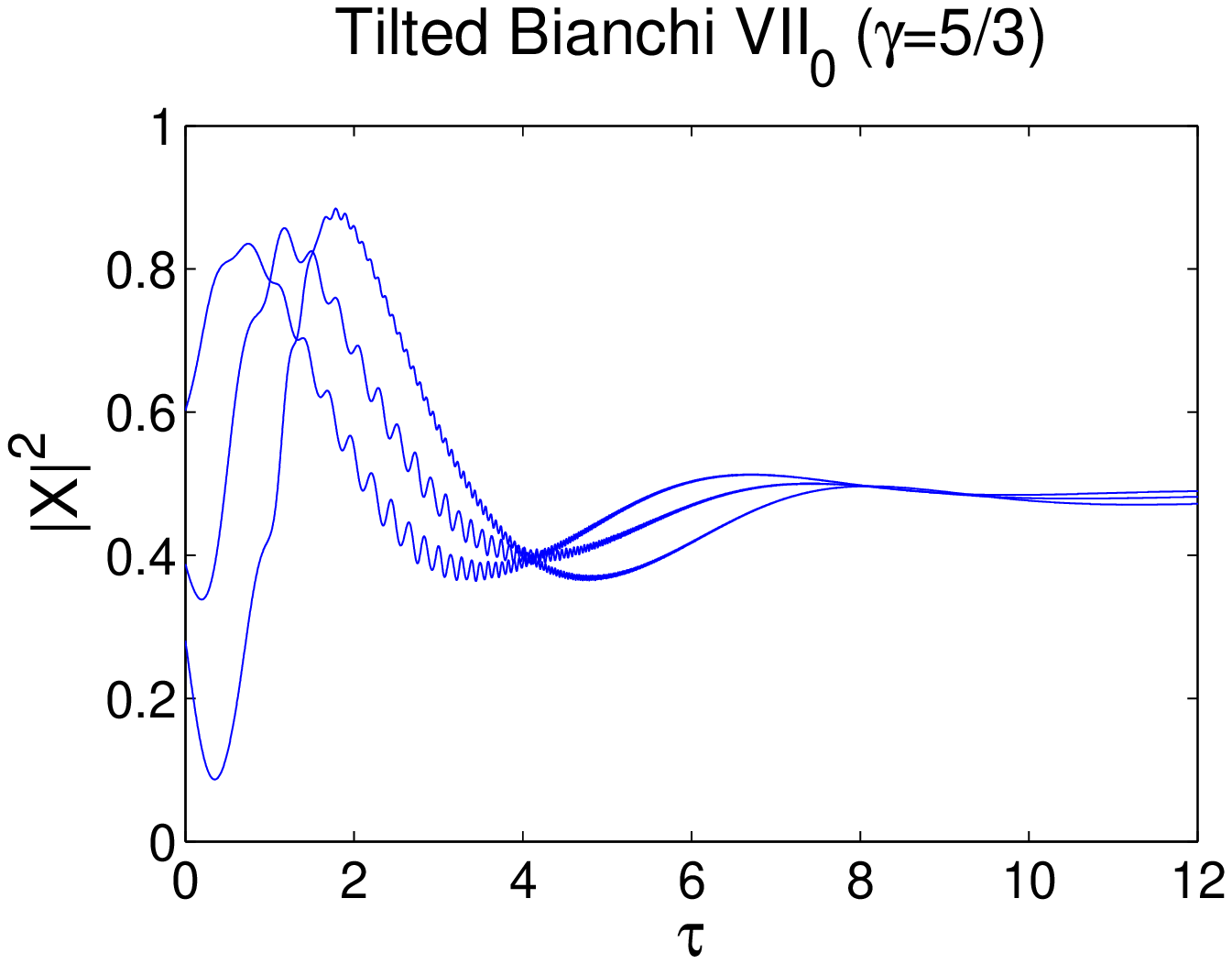}
\caption{Plots of $M$, $\Sigma_+$, $v_1$ and $|{\bf X}|^2$ for
$\gamma=5/3$, illustrating the rate at which $M$ tends to zero.
 Note how the variables $v_1$ and $|{\bf X}|^2$ do not approach a unique value.
 Also notice the oscillatory behavior and how it dampens.}\label{Fig:g=53}
 \end{figure}

\begin{figure}
\includegraphics[width=5.5cm]{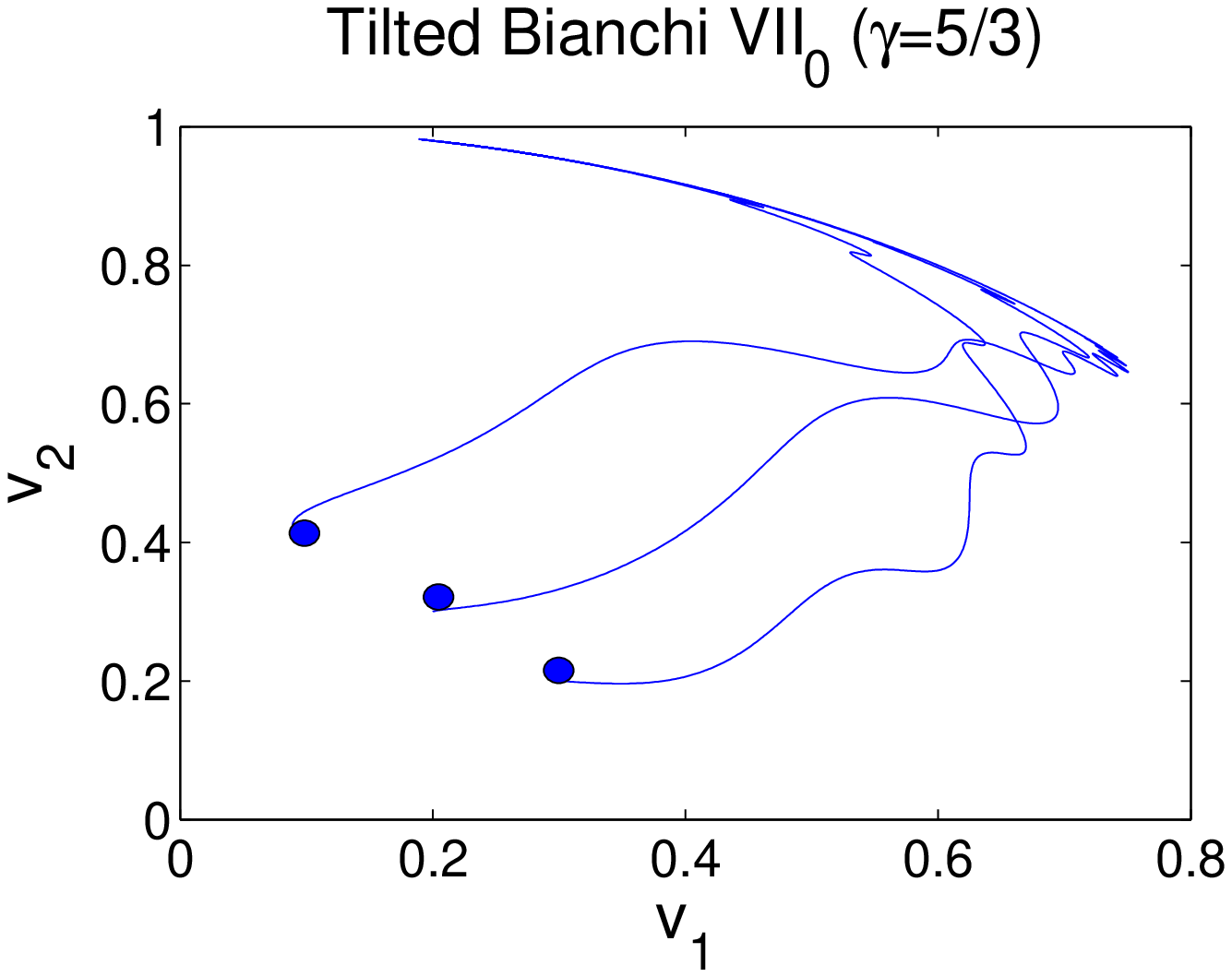}\includegraphics[width=5.5cm]{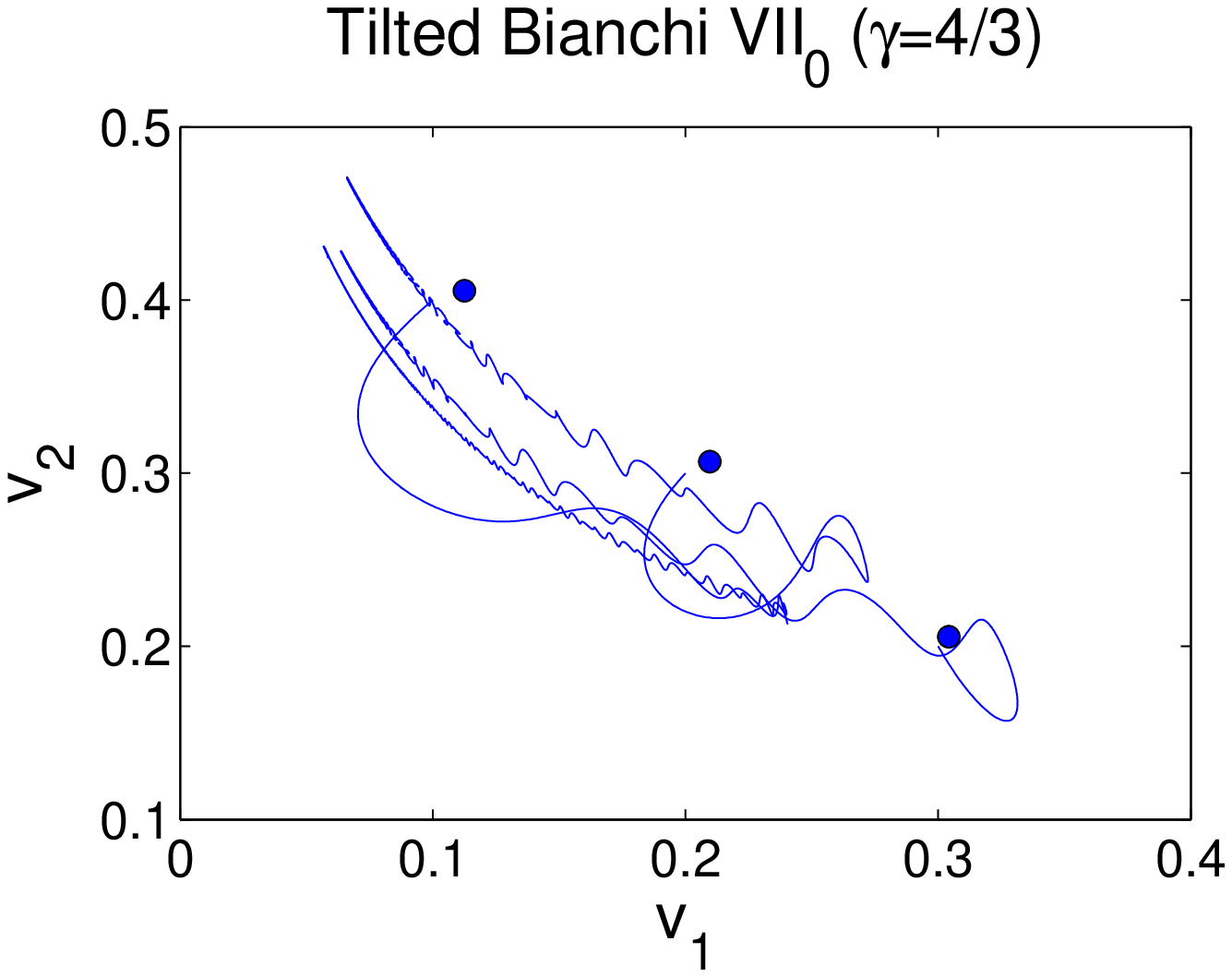}

\includegraphics[width=5.5cm]{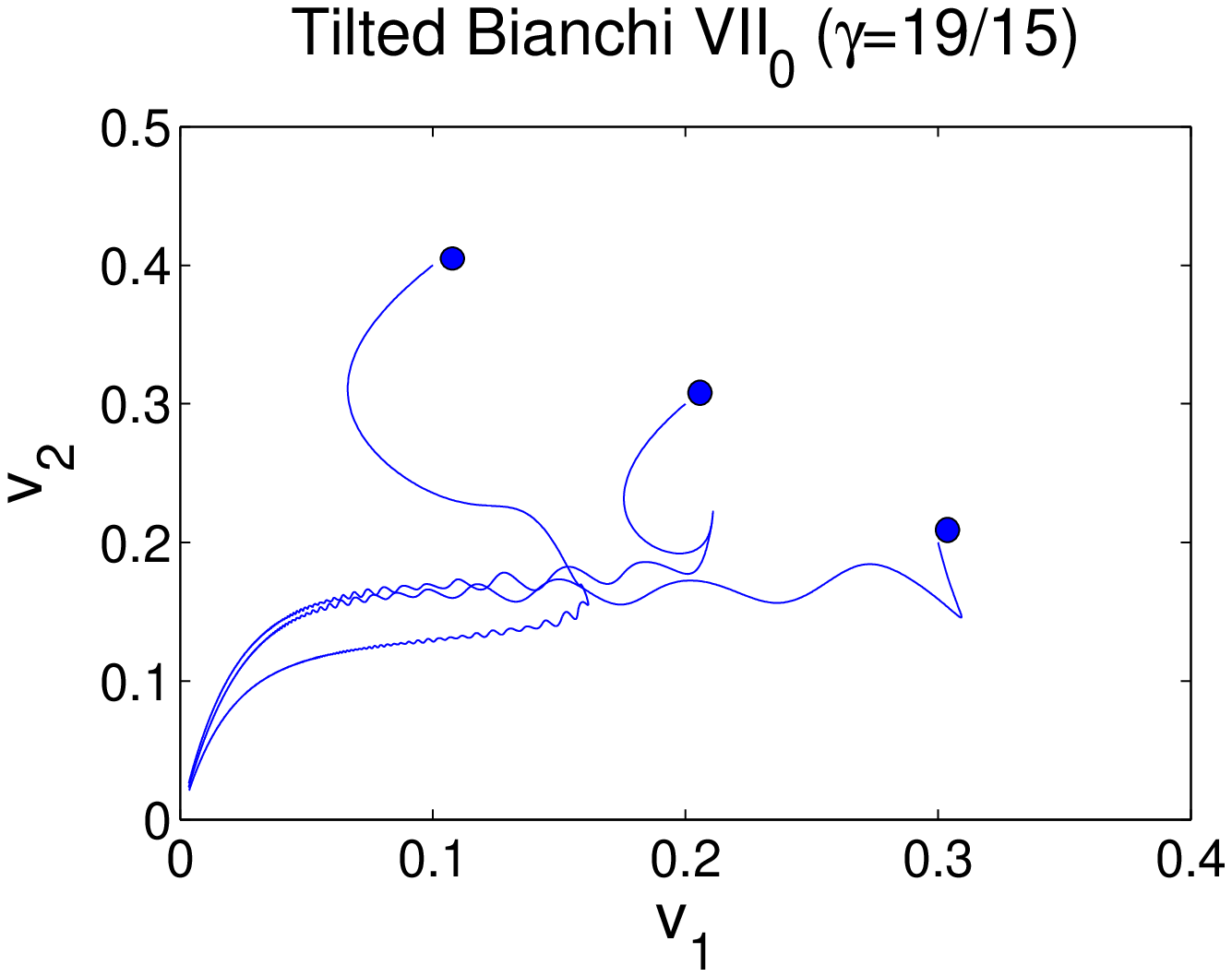}\includegraphics[width=5.5cm]{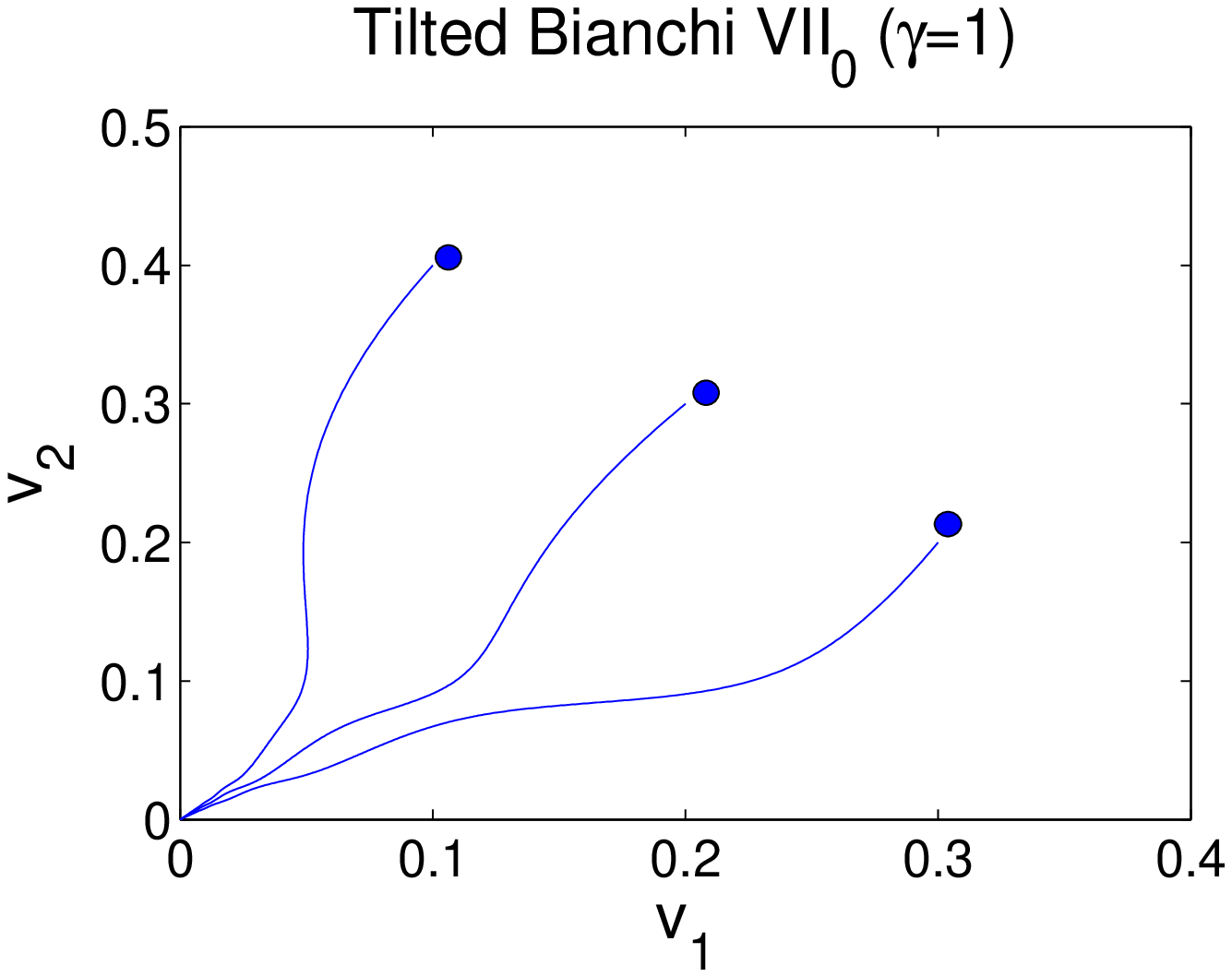}
\caption{Phase portraits of $v_1$ versus $v_2$ for different values
of $\gamma$ illustrating the fundamental differences in the
asymptotic states in each case. Note how $v_1^2+v_2^2\to 0$ for
$\gamma=1$ and $\gamma=19/15$, $v_1^2+v_2^2\to V_0^2 \not = 0$ for
$\gamma=4/3$ and $v_1^2+v_2^2\to 1$ for $\gamma=5/3$.
%The dots represent the initial value.
}\label{Fig:vs}
 \end{figure}

Many different sets of initial conditions were tested to confirm the
results of the analytical calculations.  We observed that
$M\to 0$ in every numerical run for values of $2/3<\gamma<2$.  We
also observed that the final asymptotic state obtained in the numerical
calculations confirmed what was found in the analytical
calculations. In Figures \ref{Fig:g=1}-\ref{Fig:vs}, we show some of the numerical runs for 
different representative values of $\gamma$. The value $\gamma=19/15$ is
chosen since it represents a typical value between the `special'
value of $\gamma=6/5$ and the bifurcation value $\gamma=4/3$.

\newpage
\bibliographystyle{amsplain}

\begin{thebibliography}{99}


\bibitem{EM} G.F.R. Ellis and M.A.H. MacCallum, \textit{Comm. Math. Phys.}
\textbf{12} (1969) 108

\bibitem{DS1} C.G. Hewitt and J. Wainwright in \textit{Dynamical Systems
in Cosmology}, eds: J. Wainwright and G.F.R. Ellis, Cambridge
University Press (1997)

\bibitem{DS2} A.A. Coley, \textit{Dynamical Systems
and Cosmology}, Kluwer, Academic Publishers (2003) 

\bibitem{BS} J.D. Barrow and D.H. Sonoda, \textit{Phys. Reports} \textbf{139}
(1986) 1


\bibitem{rosjan} K.Rosquist  and  R.T. Jantzen,  \textit{Phys. Reports} {\bf 166} (1988) 89%-124.

\bibitem{BN} O.I. Bogoyavlenskii
\textit{Methods in the Qualitative Theory of Dynamical Systems in
Astrophysics and Gas Dynamics} Springer-Verlag (1985).


\bibitem{KingEllis} A.R. King and G.F.R. Ellis, \textit{Commun. Math. Phys.}
\textbf{31} (1973) 209

\bibitem{HBC} D. Hobill, A.B. Burd and A.A. Coley, eds., 1994, \textit{Deterministic chaos
in general relativity}, NATO ASI Series B, vol 332 (Plenum Press, New York).

%\bibitem{Ringstrom1} H. Ringstr{\"o}m, \textit{Class. Quant. Grav.} \textbf{17} (2000) 713

\bibitem{Ringstrom2} H. Ringstr{\"o}m, \textit{Annales Henri Poincare} \textbf{2} (2001) 405


\bibitem{HWClassB} C.G. Hewitt and J. Wainwright, \textit{Class. Quant. Grav.} \textbf{10} (1993) 99

\bibitem{HBWII} C.G. Hewitt, R. Bridson, J. Wainwright, \textit{Gen. Rel. Grav.%
} \textbf{33} (2001) 65



\bibitem{Shikin} I.S. Shikin, \textit{Sov. Phys. JETP} \textbf{41} (1976) 794

\bibitem{Collins} C.B. Collins, \textit{Comm. Math. Phys.} \textbf{39}
(1974) 131

\bibitem{CollinsEllis} C.B. Collins and  G. F. R.Ellis, \textit{Phys. Rep.}
{\bf 56} (1979) 65%-105.


\bibitem{HWV} C.G. Hewitt and J. Wainwright, \textit{Phys. Rev.} \textbf{D46}
(1992) 4242

\bibitem{Harnett} 
D. Harnett, \textit{Tilted Bianchi type V cosmologies with vorticity}, Master's thesis, University of Waterloo, Canada, 1996



\bibitem{hervik} S. Hervik, \textit{Class. Quantum Grav.} \textbf{21} (2004) 2301

\bibitem{coleyhervik} A.A. Coley and S. Hervik,  \textit{Class. Quantum Grav.} \textbf{21} (2004) 4193

\bibitem{CH} A.A. Coley and S. Hervik,  \textit{Class. Quantum Grav.} \textbf{22} (2005) 579

\bibitem{HHC} S. Hervik, R.J. van den Hoogen and A.A. Coley, \textit{Class. Quantum Grav.} \textbf{22} (2005) 607


\bibitem{VII0}
J. Wainwright, M.J. Hancock and C. Uggla, \textit{Class. Quant.
Grav.} \textbf{16} (1999) 2577

\bibitem{VII0rad}
U.S. Nilsson, M.J. Hancock and J. Wainwright, \textit{Class.
Quant. Grav.} \textbf{17} (2000) 3119%-3134

\bibitem{CC1}
B.J. Carr and A.A. Coley, \textit{Class. Quant. Grav.} \textbf{16} (1999) R31

\bibitem{CC2}
B.J. Carr and A.A. Coley, \texttt{gr-qc/0508039}

%\bibitem{WCThesis}
%W.C. Lim, \textit{The Dynamics of Inhomogeneous Cosmologies}, PhD Thesis, University of Waterloo, Canada, 2004

\bibitem{LDW}
W.C. Lim, R.J. Deeley and J. Wainwright, in preparation.

\bibitem{BarrowTipler}
J.D. Barrow and F.J. Tipler, \textit{Nature} \textbf{276} (1978) 453

\bibitem{BHWeyl}
J.D. Barrow and S. Hervik, \textit{Class. Quantum Grav.}
\textbf{19} (2002) 5173

\bibitem{BHtilted} J.D. Barrow and S. Hervik, \textit{Class. Quantum
  Grav.} \textbf{20} (2003) 2841
  
  
\bibitem{Apo} P. Apostolopoulos, \texttt{gr-qc/0407040}

  
  \bibitem{BKL}
V.A. Belinskii, I.M. Khalatnikov and E.M. Lifshitz, \textit{Adv. Phys.} \textbf{31} (1982) 639


\bibitem{Wald} R.M. Wald, \textit{Phys. Rev. } \textbf{D28} (1983) 2118

\bibitem{Rendall}
A.D. Rendall, \textit{Math. Proc. Camb. Phil. Soc.} \textbf{118} (1995) 511.


\end{thebibliography}

\end{document}